\documentclass[dvipdfmx]{jpsj3}
\usepackage{txfonts,amsmath,amssymb,bm}

\title{
The Pitman-Yor process and
an empirical study of choice behavior
}

\author{
  Masato Hisakado$^1$\thanks{hisakadom@yahoo.co.jp}, Fumiaki Sano$^2$\thanks{ms16821@st.kitasato-u.ac.jp}, and
  Shintaro Mori$^3$\thanks{shintaro.mori@gmail.com}}
\inst{$^1$Nomura Holdings, Inc.,Otemachi 2-2-2, Chiyoda-ku, Tokyo 100-8130, Japan \\
  $^2$Department of Physics, Graduate
   School of Science, 
   Kitasato University \\
Kitasato 1-15-1, Sagamihara, Kanagawa 252-0373, Japan
  \\
  $^3$
Department of Mathematics and Physics,
Graduate School of Science and Technology, 
Hirosaki University \\
Bunkyo-cho 3, Hirosaki, Aomori 036-8561, Japan
} 

\abst{
This study
discusses choice behavior using a voting model in which
voters can obtain information from a finite number of previous $r$  voters.
Voters vote for a candidate with a probability 
proportional to the previous vote ratio, which is visible to the voters.
We obtain
the Pitman sampling formula as the equilibrium distribution of $r$ votes.
We present the model as a process of posting 
on a bulletin board system, 2ch.net, where users can choose one of 
many threads to create a post. We
explore how this choice depends on the last 
$r$ posts and the distribution of these last $r$ posts across threads.
We conclude that the posting process is 
described by our voting model with analog herders for a small $r$, which might
correspond to the time horizon of users' responses.
}

\begin{document}
\maketitle

\section{\label{sec:introdunction}Introduction}
In physics, equilibrium states are comparatively well understood,
whereas non-equilibrium states continue to attract much attention
\cite{Privman:1997,Hinrichsen:2000,Mantegna:2008}.
The latter states pose several interesting problems, and 
clarifying and classifying the nature of non-equilibrium
stationary states continues to be a central research theme
\cite{Sasamoto:2010}.
In other disciplines, the non-equilibrium
 stationary state is referred as the equilibrium state.
The ecology literature highlights that the equilibrium state in a
zero sum model, in which the total number of individuals is constant, is an important process \cite{Hubbell:2001}. 
The economics literature discusses the equilibrium state in which companies survive
competitive conditions \cite{Aoki:2002,Fujiwara:2004}.

The Ewens sampling formula is a one-parameter
probability distribution on the set of all partitions of an
integer \cite{Ewens:1990}. The Pitman sampling formula is a
two-parameter extension of the Ewens sampling formula \cite{Pitman:2006}.
The Pitman-Yur process \cite{Pitman:1997}
and a generalized P\'{o}lya urn 
\cite{Shibuya:2001} are the non-equilibrium
 stochastic processes that 
 derive the Pitman sampling formula \cite{Pitman:2006}.
 These processes permit new entries of individuals and an increasing number of
 them.
In a similar non-equilibrium process in which 
the number of species or vertices increases,
a power-law distribution can be obtained
\cite{Yule:1925,Simon:1955,Barabasi:1999}.

We introduced a sequential voting model in previous studies 
\cite{Hisakado:2015}. At each time step
$t$, one voter
chooses one of two candidates. In addition, the $t$th
voter can see the previous $r$ votes, and, thus, is given access to public perception.
When the voters vote for a candidate with a probability
that is proportional to the previous referable votes and
there are two candidates,
the model can be considered as Kirman's ant colony model \cite{Kirman:1993}.
In these previous studies, a beta-binomial distribution was derived as the
equilibrium distribution of the $r$ referable votes
in the stationary state of the process \cite{Hisakado:2006}.
If we assume that voters can refer to all votes, the process becomes
a non-equilibrium process. 
The equilibrium distribution
and the probability distribution in the non-equilibrium process
are the same \cite{Hisakado:2015}.
 
The response function is important for opinion dynamics, and decision-making depends on social influence. In this study, we consider the case of analog herders who vote for a candidate with a probability that is proportional to the referable votes. We refer to the response function in this case as an analog type.

On the other hand, threshold rules have been used to influence response functions in a variety of relevant theoretical scenarios 
\cite{Galam:2004, Galam:2008, Galam:2005,Galam:2005-2, Hisakado:2011}. This rule posits that individuals will choose one of two choices only when a sufficient number of other individuals have adopted that choice. We refer to such individuals as digital herders.
 From our experiments, we observe that people's individual behavior falls between that of digital herders and that of analog herders. In this study, we show that people behave as analog herders when posting to a bulletin board system.

We extend Kirman's ant
colony model when the number of candidates is greater than two and 
not fixed \cite{Kirman:1993,Hisakado:2015}.
The model is a finite reference version of the Pitman-Yor process and the generalized
P\'{o}lya urn model\cite{Pitman:1997, Shibuya:2001}.
We derive the Pitman sampling formula as an equilibrium distribution.
As a comprehensive example of the model, we analyze time series data for 
posts on 2ch.net and electoral data for 
the Japanese House of Representatives.
In the former case, votes and candidates in the voting model
correspond to posts on bulletin boards
and the bulletin boards' threads.
When $r$ is small, the posting process is described by
the voting model.
For the parliament election data, the number of candidates
is fixed. Using the Pitman sampling formula,  
we compare the correlation
between votes before and after
the introduction of the small constituency system in 1993.

  
The remainder of this paper is organized as follows.
Sec.~\ref{sec:model}
introduces a voting model, and 
we derive the Pitman sampling formula as an equilibrium distribution of votes.
Sec.~\ref{sec:four} then presents the characteristics
of several parameters.
Sec.~\ref{sec:data}
studies time series data for posts on 2ch.net using the voting model, and Sec.~\ref{sec:con} concludes. 
Appendix \ref{A} studies the case in which the number of candidates is fixed.
 Appendix \ref{C} examines Japanese election data
 as a case in which the number of candidates is fixed. Appendix \ref{B}
 and Appendix \ref{aD} provide information on the 2ch.net data and supplementary
 results of the data analysis.

\section{\label{sec:model} Model}
We examine choice behavior using a voting model with candidates
$C_1,C_2,\cdots $. At time $t$,
candidate $C_j$ has $c_j(t)$ votes.
At each time step, a voter votes for one candidate, and the voting is sequential.
Thus, at time $t$, the $t$th voter votes, after which the total number of votes is $t$.
Voters are allowed to see the 
$r$ previous votes for each candidate, where $r$ is a constant, and, thus, voters are aware of public perception.
The candidates are allowed to both enter and exit.
The voter votes for the new candidate $C_i$ with probability
$(\theta+K_r \alpha)/(\theta+r)$, where $r$ is the number of referred votes and 
$K_r$ is the number of candidates who have more than one vote in the last $r$ votes.
$\alpha$ and $\theta$ are parameters.
If a candidate does not have more than one vote in the last $r$ votes, he/she exits. 
$i$ in $C_{i}$ is the number of candidates who have appeared in the past plus one.
The number of candidates at $t=1$ is one, and there is only one candidate $C_{1}$.

In terms of the Chinese restaurant process or Hoppe's urn process,
 we describe the voting process as follows \cite{Pitman:2006, Hoppe:1984, Shibuya:2001}.
At first, there is an urn with $\theta$ black balls in it.
In each step, one ball is drawn from the urn and
two balls are placed back into the urn.
In the first turn, a black ball is drawn, and a ball of color 1 and the black ball
are placed back into the urn.
In subsequent turns, if the drawn ball is black, a ball of another color that has not
appeared in the past and the black ball are returned to the urn, and if the drawn ball is not black,
 the ball is duplicated, and the two balls are placed back into the urn. 
The difference between this voting model and the Chinese restaurant
process is that the voter refers to only the recently added $r$ balls and the black balls.

\begin{figure}[h]
\includegraphics[width=8cm]{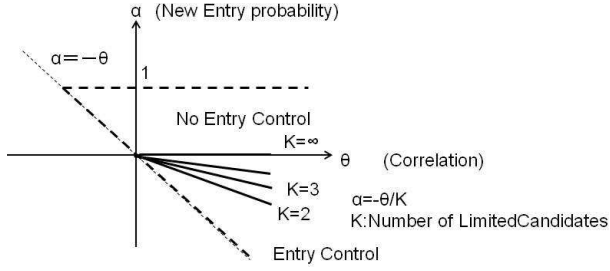}
\caption{Parameter space $\alpha$ and $\theta$.
  $\alpha$ is the parameter that adjusts the entry
  probability of new candidates as per the number of
  candidates, $K_{r}$.
  $\theta$ is the parameter that controls the overall
  probability of adding a new candidate.
  The dotted line is not included in the parameter space.
  When $\alpha<0$, the constraint $\theta=\alpha K$ exists, 
  and $K_{r}$ cannot exceed $K$.}
\label{fig:model}
\end{figure}

We illustrate the parameter space in Fig.\ref{fig:model}.
$\theta$ is the parameter that controls the overall
 probability of adding a new candidate, and
$\alpha$ is the parameter that adjusts the entry probability of
 new candidates according to the number of candidates $K_{r}=K$.
When $\alpha<0$, the constraint $\theta=-\alpha K$ exists.

We consider the case in which voters are analog herders.
If $c_j (t)\geq1$, the transition is 
\begin{equation}
c_j(t)=k \rightarrow k+1:
 P_{j,k,t:l,t-r}=\frac{-\alpha+(k-l)}{\theta+ r},
\label{pda2}
\end{equation}
where $P_{j,k,t:l,t-r}$s are the probabilities of the process.
The number of votes for $C_j$ at $(t-r)$ is $c_j (t-r)=l$.
Hence, if $(k-l)=0$, candidate $C_j$ exits the system.

The process of a new candidate $C_{j}$ entering is 
\begin{equation}
c_j(t)=0 \rightarrow 1:
 P_{j,k,t:l,t-r}=\frac{K_r \alpha+\theta}{\theta+ r},
\label{pda3}
\end{equation}
where the number of candidates who have more than $0$ votes is $K_r$. 

When $\alpha\geq 0$, from (\ref{pda2}) and (\ref{pda3}), 
the constraints $\theta+\alpha>0$ and $1>\alpha\geq 0$ exist.
(See Fig.\ref{fig:model}.)
There is no upper limit on the number of candidates.
When $\alpha>0$, the probability of a new entry increases with an increase in $K_r$.
When $\alpha=0$, the probability of a new entry is constant.
When $\alpha<0$, from ($\ref{pda3}$), the constraint $\alpha K+\theta=0$ exists.
The probability of a new entry decreases with an increase in $K_r$,
and $K$ is the upper limit of $K_r$. The number of candidates with more than
one vote does not exceed $K$.
This model is similar to the model that does not allow candidate entry, which is further discussed in Appendix \ref{A}. 

The distribution of $c_j (t)$ as the partition of integer $t$ follows the Pitman 
sampling formula in the generalized P\'olya urn
model \cite{Pitman:2006,Shibuya:2001}.
This is a non-equilibrium process, and the number of votes increases.
We focus not on the snapshot $c_j (t)$ but on the time series of state $c_j (t)-c_j(t-r)$.
This is an equilibrium process, and the number of total votes is constant.

We consider a hopping rate among $(r+1)$ states $\hat{k}_j=k-l$, $\hat{k}_j=0,1,\cdots, r$, and, here, we focus on the state.
At each $t$, the vote at time $(t-r)$ is deleted, and a new one is obtained.  
$\hat{k}_j$ is the number of votes that candidate $C_j$ obtained in the previous $r$ votes.

First, we consider the case $\hat{k}_j>1$.
The transition is
\begin{eqnarray}
\hat{k}_j&\to& \hat{k}_j+1:
 P_{\hat{k}_j,\hat{k}_j+1,t}=\frac{r-\hat{k}_j}{r}\frac{-\alpha+ \hat{k}_j}{\theta+ r-1},
\nonumber \\
\hat{k}_j &\to& \hat{k}_j-1:
P_{\hat{k}_j,\hat{k_j}-1,t}=\frac{\hat{k}_j}{r}\frac{(\theta+\alpha)+ (r-\hat{k}_j-1)}{\theta+ r-1},
\nonumber \\
\hat{k}_j &\to& \hat{k}_j:
P_{\hat{k}_j,\hat{k}_j,t}=1-P_{\hat{k},\hat{k}-1,t}-P_{\hat{k},\hat{k}+1,t}.
\label{pd1}
\end{eqnarray}
$P_{\hat{k}_j,\hat{k}_j\pm 1,t}$ and $P_{\hat{k}_j,\hat{k}_j,t}$
are the probabilities of the process. 
$P_{\hat{k}_j,\hat{k}_j\pm 1,t}$ is the product of
the probabilities of exit and entry.

We consider hopping from candidate $C_{i}$ to $C_{j}$.
\begin{eqnarray}
\hat{k}_i&\to& \hat{k}_i-1,\hat{k}_j \to \hat{k}_j+1:
 P_{\hat{k}_i \to \hat{k}_i-1,\hat{k}_j \to \hat{k}_j+1,t}=\frac{\hat{k}_i}{r}\frac{-\alpha+ \hat{k}_j}{\theta+ r-1},
\nonumber \\
\hat{k}_i&-&1 \to  \hat{k}_i,\hat{k}_j +1\to \hat{k}_j:
P_{\hat{k}_i-1 \to \hat{k}_i,\hat{k}_j+1 \to \hat{k}_j,t}=\frac{\hat{k}_j+1}{r}\frac{-\alpha+ \hat{k}_i-1}{\theta+ r-1}.
\nonumber \\
\label{pd2}
\end{eqnarray}
Here, we define $\mu_{r}(\hat{k},t)$ as the distribution function of state $\hat{k}$ at time $t$.
The number of all states is $(r+1)$. Using the fact that the process is reversible,
in the equilibrium, we have
\begin{equation}
\frac{\mu_{r}(\hat{k}_i,\hat{k}_j, t)}{\mu_{r}(\hat{k}_i-1,\hat{k}_j+1, t)}
=\frac{\hat{k}_j+1}{\hat{k}_i}\frac{-\alpha+\hat{k}_i-1}{-\alpha+\hat{k}_j}.
\end{equation}
We separate indexes $i$ and $j$ and obtain
\begin{eqnarray}
\frac{\mu_{r}^i(\hat{k}_i, t)}{\mu_{r}^i(\hat{k}_i-1, t)}
&=&\frac{-\alpha+\hat{k}_i-1}{\hat{k}_i}c,
\nonumber \\
\frac{\mu_{r}^j(\hat{k}_j+1, t)}{\mu_{r}^j(\hat{k}_j, t)}
&=&\frac{-\alpha+\hat{k}_j}{\hat{k}_j+1}c,
\label{ec}
\end{eqnarray}
where $c$ is a constant.

In the equilibrium, the number of candidates with $\hat{k}_j >0$ is $K_r$.
We ignore candidates with $ \hat{k}_j=0$ and change the number of candidates and votes
from $C_j$, $\hat{k}_j$ to $\tilde{C}_m$, $\tilde{k}_m$, where $m=1,\cdots,K_r$, wherein $\tilde{k}_m>0$.  

We can write the distribution as
\begin{equation}
\mu_r(\bm{\tilde{a}}
,\infty)=
\left(
\begin{array}{r}
\theta+r-1 \\
r
\end{array}
\right)^{-1}
\prod_{m=1}^{K_r}
\frac{(1-\alpha)^{[\tilde{k}_m]}}{\tilde{k}_m !}\mu_r(\bm{\tilde a}=\underbrace{(1,1,\cdots,1)}_{K_{r}}
,\infty),
\label{bin2}
\end{equation}
where $\bm{\tilde{a}}=(\tilde{k}_1,\cdots,\tilde{k}_{Kr})$
and $x^{[n]}=x(x+1)\cdots (x+n-1)$, which is the Pochhammer symbol.

Given (\ref{ec}),
 we can obtain the equilibrium condition between  $\bm{\tilde{a}}=\underbrace{(1,1,\cdots,1)}_{K_{r}}=
\bm{1}$ and $\bm{\tilde{a}}=\underbrace{(0,0,\cdots,0)}_{K_{r}}=\bm{0}$:
\begin{equation}
\frac{\mu_{r}^i(\bm{\tilde{a}}=(\overbrace{1,\cdots,1}^{n},0,\cdots,0), \infty)}{\mu_{r}^i(\bm{\tilde{a}}=(\underbrace{1,\cdots,1}_{n-1},0,\cdots,0), \infty)}
=\frac{\theta+(n-1)\alpha}{n}c,
\end{equation}
where $n=1, \cdots,K_r$.
Hence, we can obtain
\begin{equation}
\mu_r(\bm{\tilde{a}}=\bm{1}
,\infty)=\prod_{m=1}^{K_r}\frac{\theta+(m-1)\alpha}{m}\mu_r(\bm{\tilde{a}}=\bm{0}
,\infty)=\frac{(\theta)^{[K_r:\alpha]}}{K_r !},
\end{equation}
where $x^{[n:\alpha]}=x(x+\alpha)\cdots (x+(n-1)\alpha)$.
Therefore, we can write (\ref{bin2}) as
\begin{equation}
\mu_r(\bm{\tilde{a}},\infty)=
\left(
\begin{array}{r}
\theta+r-1 \\
r
\end{array}
\right)^{-1}
\frac{\theta^{[K_r:\alpha]}}{K_r !}
\prod_{j=1}^{r}
\left(
\frac{(1-\alpha)^{[j-1]}}{j!}
\right)
^{a_{j }},
\end{equation}
where $a_{j }$ is the number of candidates who have $j$ votes.
Therefore, the number of candidates $\sum_{j=1}^{r}a_j=K_r$ and
that of votes $\sum_{j=1}^{r}j a_j=r$ are related.
Hereafter, we use a partition vector $\bm{\hat{a}}=(a_1,\cdots, a_r)$.

We consider the partitions of the integer $K_r$. 
To normalize, we add the following combination term, $K_r !/a_1 !\cdots a_r !$:
\begin{equation}
\mu_r(\bm{\hat{a}},\infty)
=
\frac{r! \theta^{[K_r:\alpha]}}{\theta^{[r]}}
\prod_{j=1}^{r}(\frac{(1-\alpha)^{[j-1]}}{j!})^{a_j}\frac{1}{a_j!}.
\label{pit2}
\end{equation}
(\ref{pit2}) is simply a Pitman sampling formula \cite{Pitman:2006}.
In the limit $\alpha=0$, we can obtain the Ewens sampling formula \cite{Ewens:1990}:
\begin{equation}
\mu_r(\bm{\hat{a}},\infty)=\frac{r! }{\theta^{[r]}}
\prod_{j=1}^{r}(\frac{\theta}{j})^{a_j}\frac{1}{a_j!}.
\label{esf}
\end{equation}
The Ewens sampling formula for the equilibrium process is presented in \cite{Aoki:2002}.

\section{\label{sec:four} Four regions in the parameter space}

In this section, we characterize four regions in the parameter space.
The parameter $\theta$ denotes the intensity of  correlations,
and $\alpha$ refers to the intensity of competition.
In the upper plane, newcomers increase
with a rise in $\alpha $.
In the lower plane, 
the probability of the minimum  growing increases
 with a decrease in $\alpha$.

\begin{figure}[h]
\begin{center}
\begin{tabular}{c}
  \includegraphics[width=8cm]{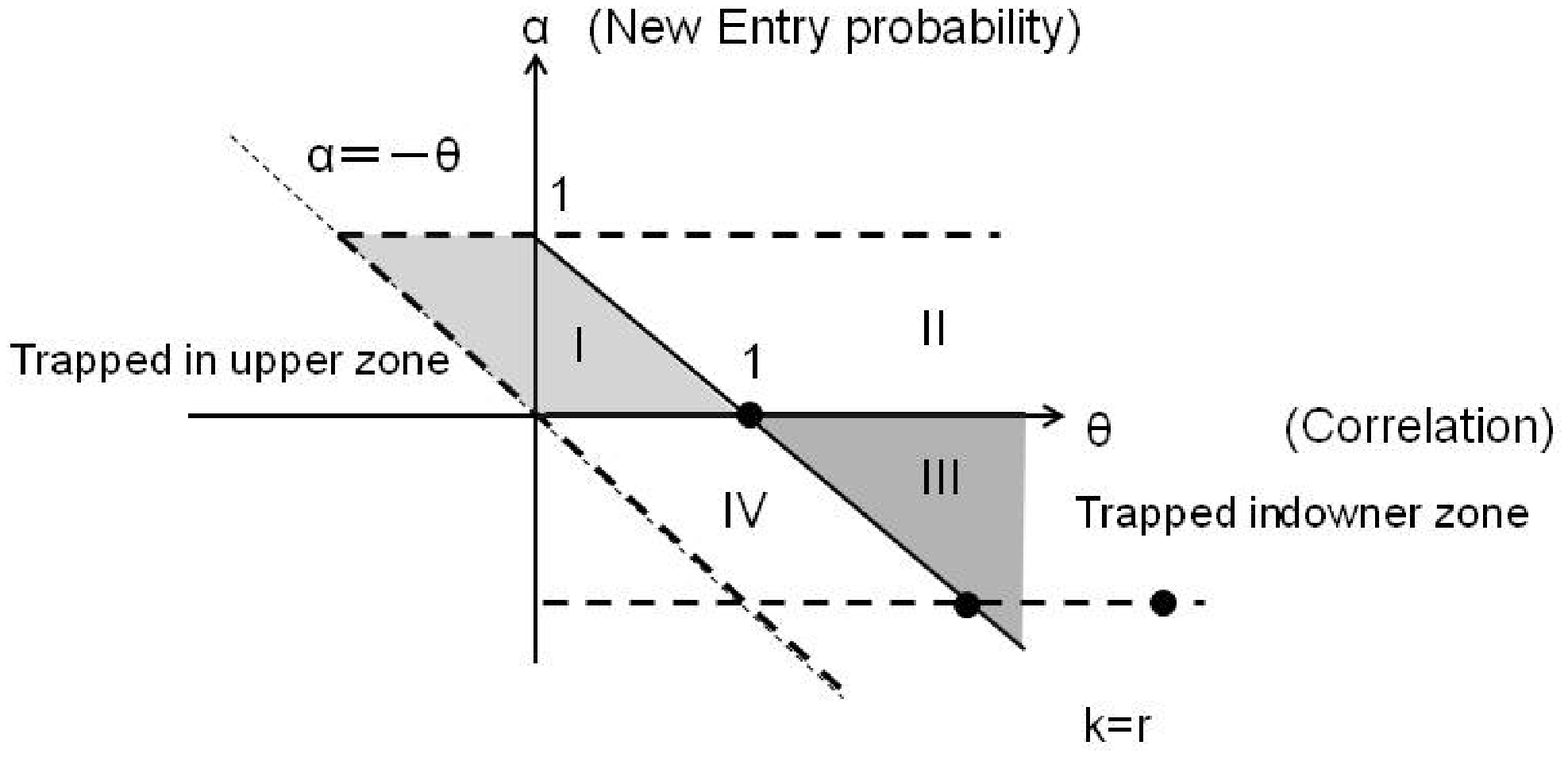} \\
\vspace*{0.2cm} \\  
\includegraphics[width=6cm]{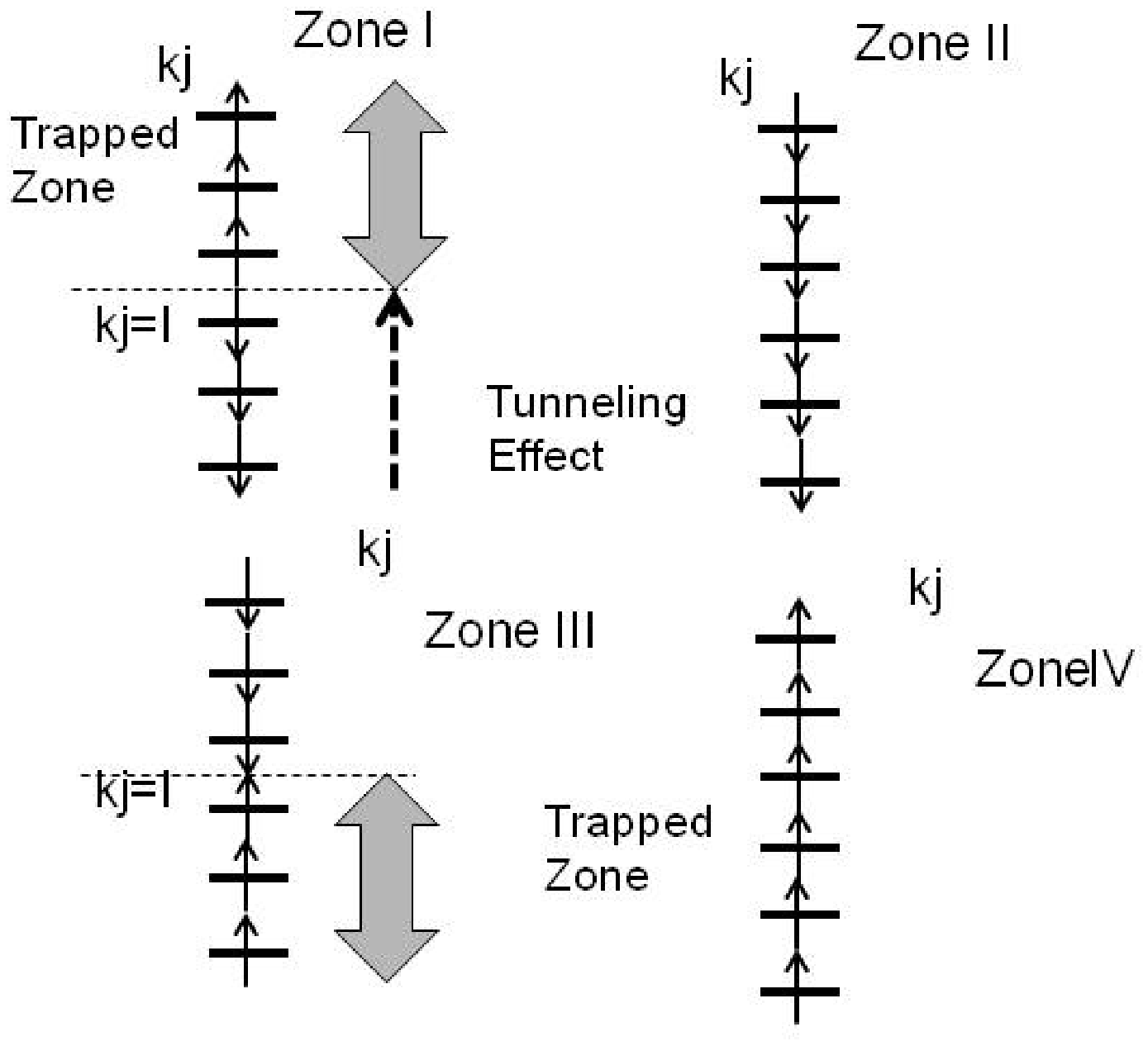} 
\end{tabular}
\caption{Four regions in the parameter space and the average increase and decrease in each region.}
\label{fig:four}
\end{center}
\end{figure}

We consider both an increase and a decrease in $\hat{k}_j$.
 The probability of a decrease in votes for candidate $j$ is $\hat{k}_j/r$, and 
that of an increase in votes for candidate $j$ is $(-\alpha+\hat{k}_j)/(\theta+r-1)$, as shown in (\ref{pda2}).
We consider the condition in which the probability of an increase is larger than that of a decrease:
\begin{equation}
\alpha\leq -\frac{\hat{k}_j}{r}(\theta-1).
\label{cond}
\end{equation}
In this region, the number of votes increases on average.

We divide the parameter space into four regions, $I 
\sim IV$, as shown in Fig. \ref{fig:four}(a).
To clarify the regions, we define zone $U_{\hat{k}_j}$ as the region
where the probability of an increase in votes is larger than that
of a decrease. $U_{\hat{k}_j}$ is defined in (\ref{cond}).
$U_{r}$ is $\alpha\leq -(\theta-1)$, and
$U_{0}$ is $\alpha\leq 0$.

We define zone I, where $\alpha>0$ and $\alpha< (1-\theta)$, as $U_r$.
In zone I, $U_{r}\supset U_{r-1}\supset \cdots \supset U_0$. 
We consider the case in which the parameter set $\bm{x}=(\theta, \alpha)$ is $\bm{x}\in U_{l}$ and 
$\bm{x}\notin U_{l-1}$.
In this case, $k_j\geq l$ is the increasing zone and $k_j< l$
 the decreasing zone. 
If a candidate has more than $l$ votes, he/she can increase the number of votes and maintain his/her position.
On the other hand, it is difficult to increase votes if the candidate has less than $l$ votes.
We show the average trend in Fig 2.(b).
In this region, the leader in the trapped zone has an advantage.
On the other hand, the competition intensifies for newcomers
as $\alpha$ increases.

We define zone II as $\alpha>0$ and $\alpha\geq (1-\theta)$.
In zone II, $\bm{x} \notin U_0,\cdots, U_r$. 
Furthermore, it is difficult to increase the number of votes for every candidate and to be a stable leader. The zone becomes more competitively intense with an increase in $\alpha$.
 In other words, it is possible to adjust the competitive intensity and protect newcomers
by adjusting $\alpha$, which denotes the number of newcomers.

In the plane in the lower half, there is a capacity limit and no newcomers.
We define zone III as $\alpha<0$ and $\alpha> (1-\theta)$.
When $\alpha<0$, $U_{0}\supset U_{1}\supset \cdots \supset U_{r}$.
We consider the case in which the parameter set $\bm{x}=(\theta, \alpha)$ is $\bm{x}\in U_{l}$ and 
$\bm{x}\notin U_{l+1}$.
In this case, $\hat{k}_j > l$ is the decreasing zone and $\hat{k}_j \leq l$
 the increasing zone.
It is easy to increase the number of votes to $\hat{k}_j=l$, but it is difficult to increase the number of votes above $\hat{k}_j=l+1$.
In this region, it is also difficult to be a stable leader.

We define zone IV as $\alpha<0$ and $\alpha\leq (1-\theta)$.
 In this zone, $\bm{x} \in U_{0},\cdots,U_r$. It is easy to increase the votes for each candidate. In addition, this zone 
is competitive when the number of members is fixed.

Next, we consider the Ewens sampling formula on the $\theta$ axis.
When $\alpha=0$ and $\theta=1$, the probabilities of an increase and decrease both become $\hat{k}_j/r$.
For any $\hat{k}_j$, the probabilities of an increase and decrease are equal.
Thus, the correlation becomes $\rho=1/2$ (see Appendix \ref{A}),
and there is a uniform random permutation.
The probability of an increase or a decrease is proportional to the number of partitions.

When $\alpha=0$ and $\theta<1$ within the boundaries of zone IV, if candidates can enter, the number of votes easily increases. In this zone, the correlation is high.
When $\alpha=0$ and $\theta>1$ within the boundaries of zone II, it is difficult to increase the number of votes. Here, there is a low correlation.
 In summary, there are numerous candidates who have few votes.

\section{\label{sec:data} Data analysis of a bulletin board system}

In this section, we examine the data of posts to a
 bulletin board system (BBS), 2ch.net. 
2ch.net is the largest BBS in Japan and
covers a wide range of topics.
Each bulletin board is separated by a field unit or a
category, such as, for example, news, food and culture, and
net relations. 
Each category is further divided into genres, or
boards, and each board contains numerous 
threads, which are segregated
by topics that belong to the board. Writing and
viewing boards is done on a thread.
There are about 900 boards on 2ch.net.
It is possible to make anonymous posts on all threads.

\begin{table*}[tbh]
\caption{Statistics of 2ch.net post data.
  The observation period, the total number of threads $N$, 
  the average number of posts per thread $T/N$, and its standard deviation (S.D.)
   are presented in the third, fourth, fifth, and sixth columns, respectively.
    $w_{Max}$ in the seventh column lists the maximum number of posts
   on a thread. The numerical value in the eighth column indicates
   the average lifetime of a thread [in days]. The lifetime is defined as
   the difference
   between the last and first post date.
   $s_{H}$ is in the last column and indicates the time horizon,
   which is defined in eq.(\ref{eq:2ch_horizon}). 
}
\label{tab:2ch}
\begin{center}
\begin{tabular}{c c c c c c c c c}
\hline
No. & Board Name  & Obs. Period & $N$ & $T/N$   & S.D. & $w_{Max}$ & Lifetime & $s_{H}$[sec] \\ 
\hline  
1 &  Business News & Aug. 10, 2009--Dec. 31, 2009& 8,248 & 140 & 290  & 7,707 &  7.5 & 260.0 \\  
2 &  East Asia News & Mar. 8, 2009--Aug.5, 2009 & 8,225 & 388 & 1,022  & 27,966 &  7.5 & 205.4\\  
3 &  Live News  & Mar. 8, 2009--Aug. 5, 2009& 15,307 & 53 & 333  & 30,443 &  2.0 & 95.1\\  
4 &  Music News & Mar. 8, 2009--Aug. 5, 2009& 23,000 & 332 & 1,123  & 78,388 &  2.8 & 140.2 \\  
5 &  Breaking News & Mar. 8, 2009--Dec.31, 2009& 33,677 & 658 & 1,497  & 113,220 &  2.9 & 94.5 \\  
\hline
6 &  Digital Camera & Aug. 10, 2009--Dec. 31, 2009& 835 & 527 & 1,530  & 33,494    &  251.4 & \\  
7 &  Game & Mar. 8, 2009--Aug. 10, 2009& 1,371 & 241 & 286  & 2,043  &  132.6 & \\  
8 &  Entertainment  & Aug. 10, 2009--Dec. 31, 2009& 1,464 & 134 & 1188  & 30999  &  35.4 & \\  
9 &  Int. Affairs & Aug. 10, 2009--Dec. 31, 2009& 688 & 233 & 241  & 1,000 &  1,123.8 & \\  
10 &  Press & Aug. 10, 2009--Dec. 31, 2009& 1,011 & 235 & 296  & 2,182 &  358.5 & \\  
\hline
\end{tabular}
\end{center}
\end{table*}

We study the time series of posts on the following ten boards:
business news, East Asia news, live news, music news,
breaking news, digital camera, game, entertainment, international affairs, and
press. We label these boards as the No. 1, No. 2, $\cdots$ No. 10 boards, respectively. 
The first five boards fall under the news category.
Each board has several hundred threads, and 
managers maintain the
number of threads by removing old ones and
replacing them with new threads. The duration of a post on a thread 
is set to five days for the No. 4 and No. 5 boards.
As a result, the lifetime of a thread 
is generally about a few days.
One cannot post more than 1,000
posts to a thread.
Threads that no longer allow posts are deleted from
the thread lists of the boards, and 
managers 
prepare a new sequential thread using the same thread title. 
The lifetime of a thread can therefore be longer than the abovementioned duration, as
 the postable duration rule applies to descendant threads with a new start date.
We identify sequential descendant threads from a common ancestor thread
as one thread.
Table \ref{tab:2ch}
summarizes the statistics of the threads of the ten boards.

The total number of posts on each board is about 0.2--22 million.
Each thread has an average of several hundred posts, and the standard deviation is large.
The maximum number of posts $w_{Max}$ is 50--100
 times larger than the average. 
The average lifetime of a thread 
on news boards is several days, which is derived from the
strict rule defining the period within which a post can be made on threads.
There is no strict rule for the remaining five boards, and
 the average lifetimes are considerably longer than those of the news boards.

\begin{figure}[tbh] 
\begin{tabular}{c}
\includegraphics[width=5cm]{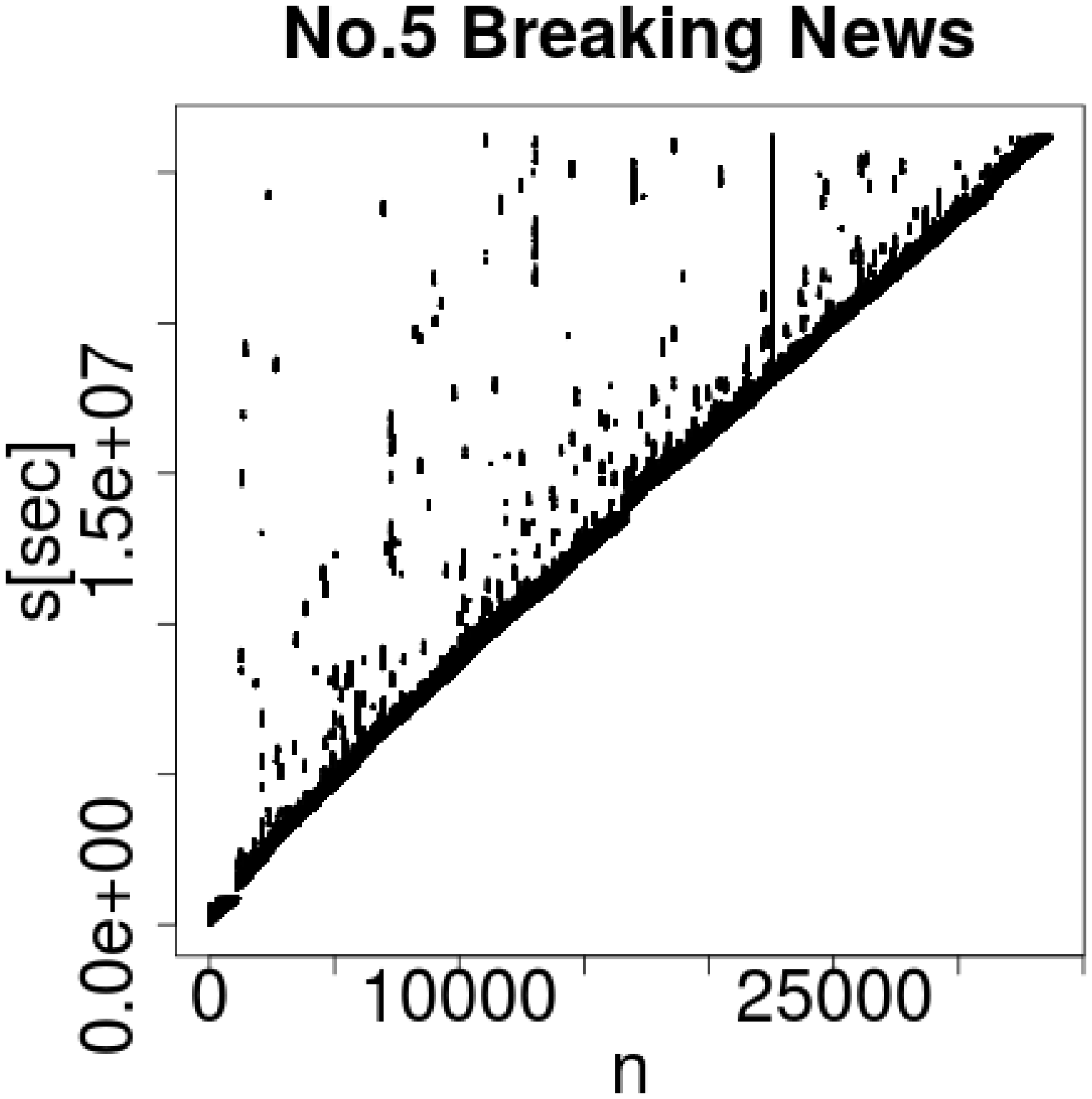} \\
\includegraphics[width=5cm]{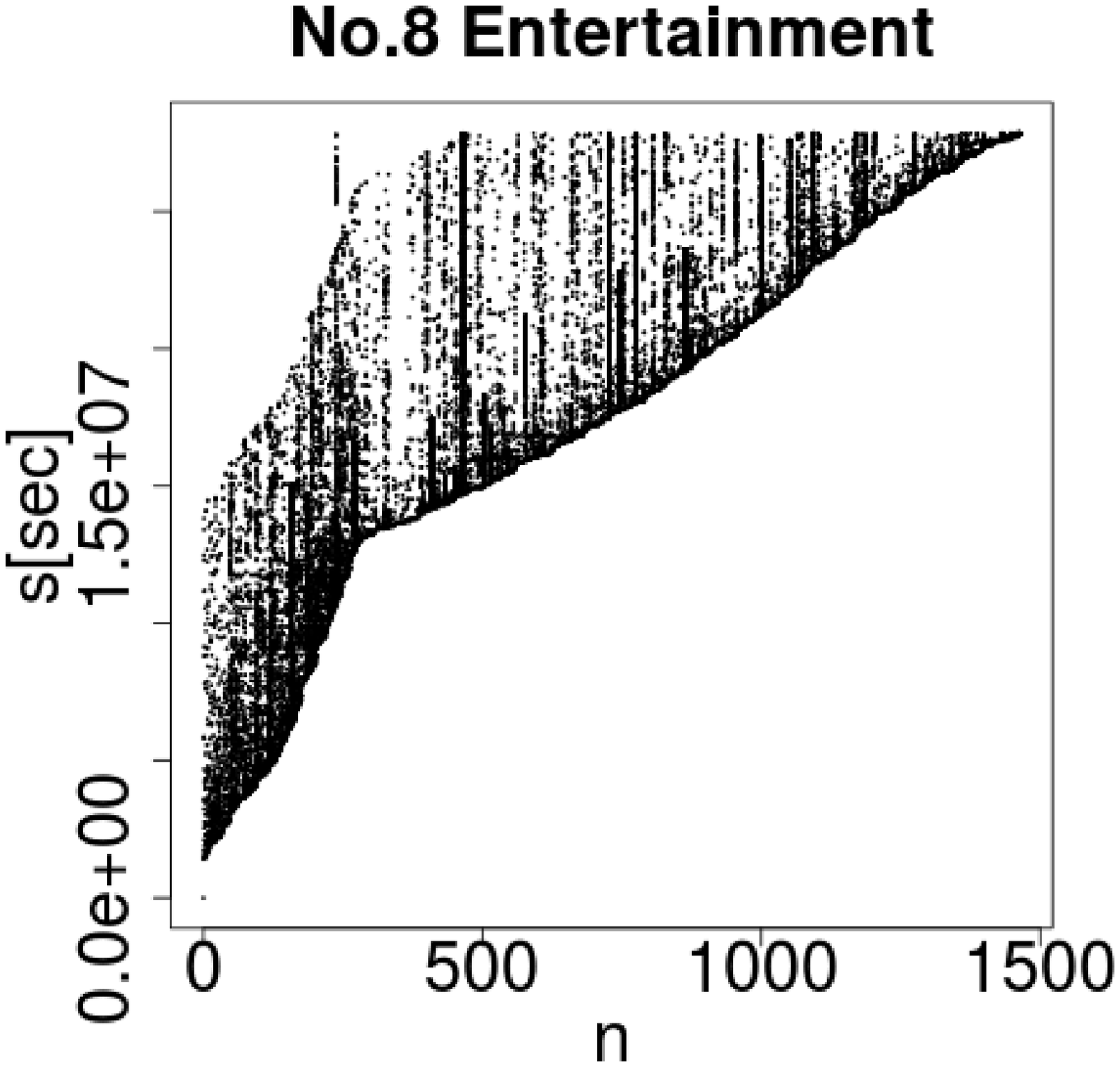} 
\end{tabular}
\caption{Scatterplot of post data $(n(t),s(t))$ for the No. 5 and No. 8 boards.
  Each dot corresponds to a post.}
\label{fig:2ch1}
\end{figure}
 
We label threads by $n\in \{1,\cdots,N\}$, and $N$ is the total number
of threads that appear on the board.
We describe the $t$th post to the board at $s[sec]$
by the thread number $n$ and $s$ as $(n(t),s(t)),t=1,\cdots,T$.
We measure the post time $s$ by setting the time of the first post time on the board as zero seconds.
We present the scatterplot of the time series post
data $(n(t),s(t)),t=1,\cdots, T$ in the $(n,s)$ plane of the No. 5 and No. 8 boards 
in Fig. \ref{fig:2ch1}.
Because the threads have a strict finite lifetime of five days on the No. 5 board,
the plot shows a narrow strip pattern. Some threads have a longer lifetime because
they have a long family tree from ancestors to descendants.
 The No. 8 board shows a wide strip pattern, which indicates that the number of threads
 is considerably large.

\subsection{Correlation function $C(\tau)$ for equilibrium $r$}

\begin{figure}[h]
\includegraphics[width=7cm]{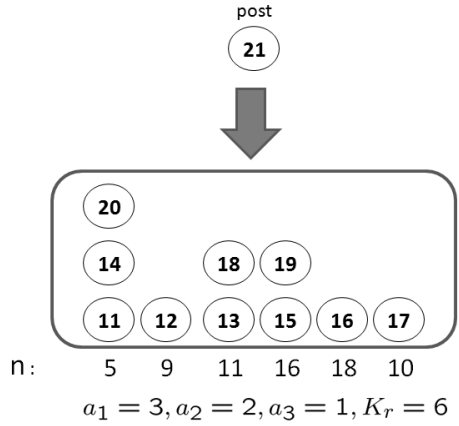}
\caption{Posts, threads, and a board.
A board is represented as a rectangle with rounded corners.
Posts are shown as balls and are labeled as $t$.
There are $K_{r}=6$ threads with a non-zero number of posts.
$n$ shows the  threads numbers. 
The "21st post? refers to the previous 10 posts, $t=11,12,\cdots.10$.
The thread numbers are $n=5,9,11,16,18,10$ and appear
$(k_{5},k_{9},k_{11},K_{16},k_{18},k_{10})=(3,1,2,2,1,1)$ times, respectively.
The multiplicities $\bm{a}$ are $a_{1}=3,a_{2}=2$ and $a_{3}=1$.
$a_{1}+a_{2}+a_{3}=K_{r}=6$ and
$1\cdot a_{1}+2\cdots a_{2}+3\cdot a_{3}=10$ hold.
The probability of a post on a thread with two posts is $2\cdot
\frac{2-\alpha}{\theta+10}$.
The probability of a thread not appearing in ten posts
is $\frac{\theta+6\cdot \alpha}{\theta+10}$.
}
\label{fig:urn}
\end{figure}
We identify threads and posts as candidates and
votes in the voting model in Fig. \ref{fig:urn}. As previously shown,
when voting occurs with reference to the previous $r$ votes,
the stationary distribution of the $r$ consecutive previous votes
obeys the Pitman sampling formula in (\ref{pit2}). 
In this case, the correlation
between $n(t)$ and $n(t-\tau)$ for the voting lag $\tau$
does not decay for $\tau< r$,
as the distribution is stationary in $r$ consecutive votes.
In the range $\tau>r$, $C(\tau)$ dumps, so if a post on the board is described by this voting model
with reference $r$, $C(\tau)$ should demonstrate
this feature.
We adopt the
expectation value of the coincidence of $n(t)$ and $n(t-\tau)$
as the correlation between $n(t)$ and $n(t-\tau)$,
\[
C(\tau)\equiv \mbox{E}(\delta_{n(t),n(t-\tau)}).
\]
We assume that $C(\tau)$ does not depend on $t$, and we estimate it using
time series data
$\{n(t)\},t=1,\cdots,T$ as
\[
C(\tau)=\frac{1}{T-\tau}\sum_{t=1}^{T-\tau}\delta_{n(t),n(t+\tau)}.
\]

\begin{figure}[tbh] 
\begin{tabular}{c}
\includegraphics[width=5.5cm]{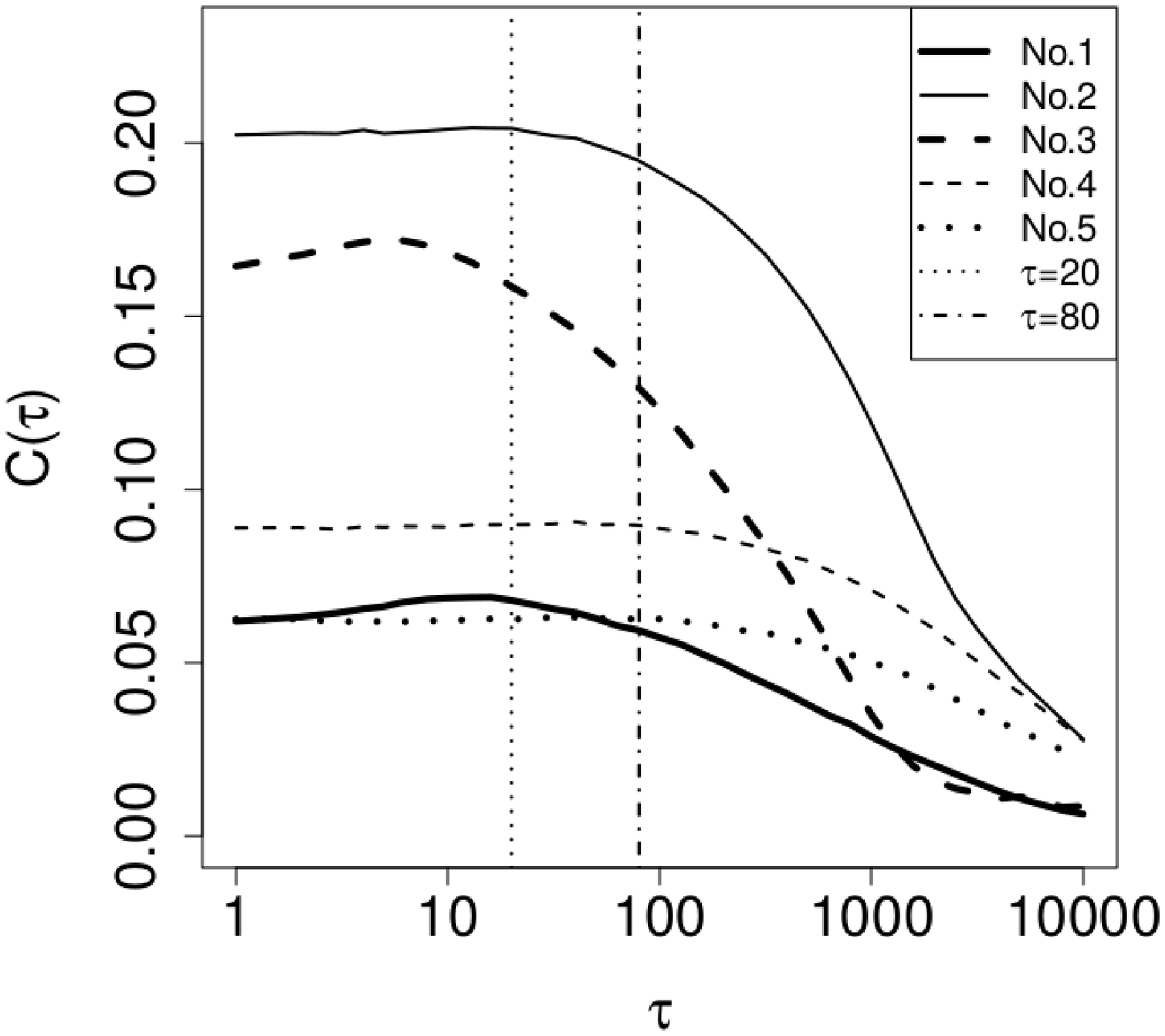} \\
\includegraphics[width=5.5cm]{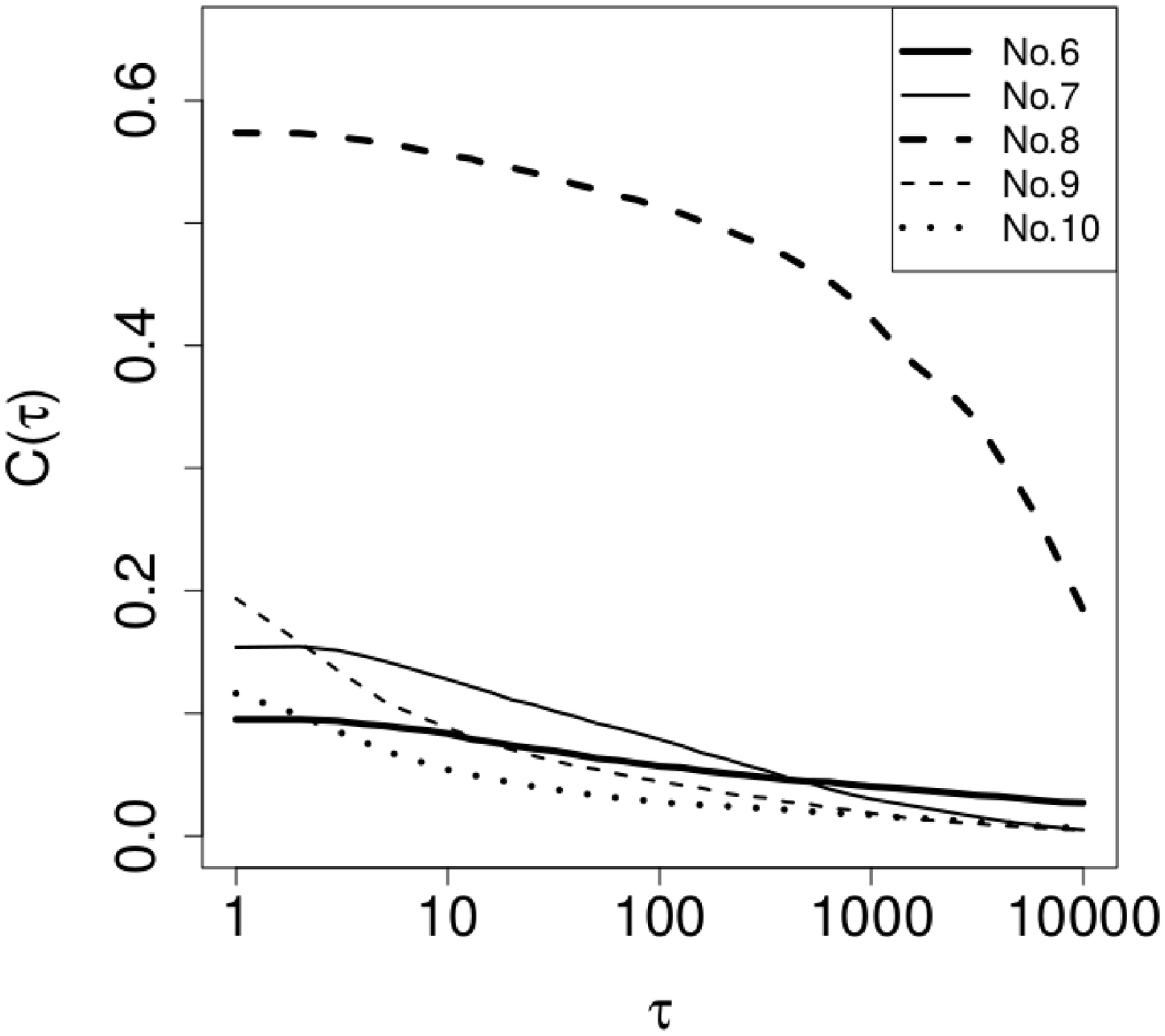} 
\end{tabular}
\caption{Plot of $C(\tau)$ vs. $\tau$.
  The left (right) panel shows the results for the first (remaining)
  five boards. $\tau$ on the x-axis indicates the voting lag and
  $C(\tau)$ denotes the auto-correlation function between $n(t)$ and
  $n(t+\tau)$.
}
\label{fig:Cor}
\end{figure}

Fig.\ref{fig:Cor}. illustrates the semi-logarithmic plot
$C(\tau)$ vs. $\tau$. 
The properties of $C(\tau)$ of a voting model with finite
$r$ are summarized in \cite{Mori:2015}.
We see a plateau structure in which $C(\tau)$ does not
decrease with $\tau$ for the first five news boards in the left panel.
For the No. 4 and No. 5 boards,
$C(\tau)$ is almost constant for $\tau\le \tau_{c}\simeq 80$.
As for the other boards, $C(\tau)$ decreases for $\tau\ge \tau_{c}=20$ for
the No. 1 and No. 2 boards
and for smaller values of $\tau_{c}\sim 5$ for the No. 3 board.
In these news boards, $C(\tau)$ is almost constant
among posts within $\tau_c$. 
On the other hand, $C(\tau)$ decreases with $\tau$ for the
latter boards in the right panel, with the exception of the No. 8 board.
As for the No. 8 board, $C(\tau)$ is large for large $\tau$ and
gradually decreases, indicating that the board has special features.

We apply the voting model to the posts on 2ch.net
in the boards for the news category with $r<\tau_{c}$.
We adopt $r=80$ for the No. 4 and No. 5 boards, $r=20$ for the No. 1 and No. 2
boards, and $r=5$ for the No. 3 board. To interpret
$\tau_{c}$, we highlight the response times of board users.
We believe that a user needs several minutes to respond to 
posts. Next, the post should
be random for a short time interval, and
the probability of a post on a thread is roughly
estimated as the post ratio in the previous posts.
In the last column of Table \ref{tab:2ch}, we indicate the time horizon $s_{H}[sec]$
for $\tau_{c}$,
which is defined as the mean duration between posts multiplied
by $\tau_{c}$, as
\begin{equation}
s_{H}\equiv \frac{s(T)-s(1)}{T-1}\cdot \tau_{c} \label{eq:2ch_horizon}.
\end{equation}
$s_{H}$ is about 1.5 to 4 minutes, which is possibly 
the requisite time duration to respond to posts.

\subsection{Estimation of the parameters $\theta and \alpha$}

We use time series data $\{n(t)\},t=1,\cdots,T$, $n\in\{1,\dots,N\}$ and 
estimate the model parameters $\theta$ and $\alpha$ using
the maximum likelihood principle. In the model, the probability of a post 
on a thread that appears in the past $r$ posts $\hat{k}$ times 
is defined as in (\ref{pda2}).
The probability of a post to $a_{\hat{k}}$ threads with $\hat{k}$ is
\begin{equation}
P_{Existing}(\hat{k})=a_{\hat{k}}\cdot \frac{\hat{k}-\alpha}{\theta+r} \label{eq:2ch_1}.
\end{equation}
The probability of a post on a new thread that does not appear in the past
$r$ posts depends on the number of threads $K_r$ in the past $r$ posts 
and is defined in (\ref{pda3}):
\begin{equation}
P_{New}(K_r)=\frac{K_r\alpha+\theta}{\theta+r} \label{eq:2ch_2}.
\end{equation}
For $t\in [1\times 10^{4},T-r]$, we choose $t_{n},n=1,\cdots S$ randomly and
study the following $r+1$ sequence, $n(t_{n}),n(t_{n}+1),\cdots,n(t_{n}+r-1)$.
We estimate the number of threads $K_r$  
and the number of threads with $\hat{k}$ posts $a_{\hat{k}}$ in the past $r$ posts.
$\sum_{\hat{k}} a_{\hat{k}}=K_r$ holds. If thread $n(t_{n}+r)$ does not
exist in the $K_r$ threads, the likelihood is $P_{New}(K_r)$.
If the thread exists and thread $n(t_{n}+r)$ appears $\hat{k}$ times,
the likelihood is $P_{Existing}(\hat{k})$.
The likelihood of $S$ sample is then estimated 
by the products of these likelihoods for all $s=1,\cdots,S$.
We adopt $S=2\times 10^{5}$.
In addition, we fit the parameters using the maximum likelihood principle
for the distribution of the
partitions of $a_{\hat{k}}$ with the Pitman sampling
formula in (\ref{pit2}).

\begin{table}[tbh]
  \caption{Fitting results of $\theta$ and $\alpha$ for 
    probabilistic rules in (\ref{eq:2ch_1}) and (\ref{eq:2ch_2}).
    We use the maximum likelihood principle, and the sample number $S$ is
    $2\times 10^{5}$.
  We show the estimates for the No. 1 and No. 2 boards with $r=20$, for the No. 3 board with $r=5$,
 and for the No. 4 and No. 5 boards with $r=80$ in the third and fourth columns.
  For $r=5$ and $r=20$, we show the goodness-of-fit results using the Pitman
  sampling formula in (\ref{pit2}) in the fifth and sixth columns.
  We adopt the same samples for the two fittings.
  The standard error (S.E.) in the last digit of the estimate is provided
  in parentheses.
  }
\label{tab:fit}
\begin{center}
\begin{tabular}{|c|c|c|c||c|c|c|}
\hline    
 &  &\multicolumn{2}{|c||}{Fit with Probabilistic Rules}&
\multicolumn{2}{|c|}{Fit with Pitman's Distribution}
\\
\hline
No. & $r$ &$\theta$(S.E.)&$\alpha$(S.E.)&$\theta$(S.E.) &$\alpha$(S.E.)\\ 
\hline  
1 & 20 & 8.8(2)  & 0.37(1) & 8.7(0)  & 0.390(2)\\  
2 & 20 & 2.2(1)  & 0.42(1) & 2.0(0)  & 0.418(2)\\  
\hline
3& 5& 1.7(0) & 0.560(4) &1.3(0) &0.623(3) \\
\hline
4 & 80 & 10.0(3)  & 0.35(1)& NA & NA   \\  
5 & 80 & 11.9(4)  & 0.28(1)& NA & NA   \\  
\hline
\end{tabular}
\end{center}
\end{table}

The estimated values for the parameters are summarized in Table
\ref{tab:fit}. 
We adopted $r=20$ for the No. 1 and No. 2 boards, $r=5$ for the No. 3 board, 
and $r=80$ for the No. 4 and No. 5 boards by the correlation analysis.
The standard errors are estimated using the square root of the negative
eigenvalue for the Hessian of the log likelihood.
For $r=80$, we only show the results by fitting with probabilistic rules.

\begin{figure}[tbhp] 
\begin{tabular}{c}
\includegraphics[width=5.5cm]{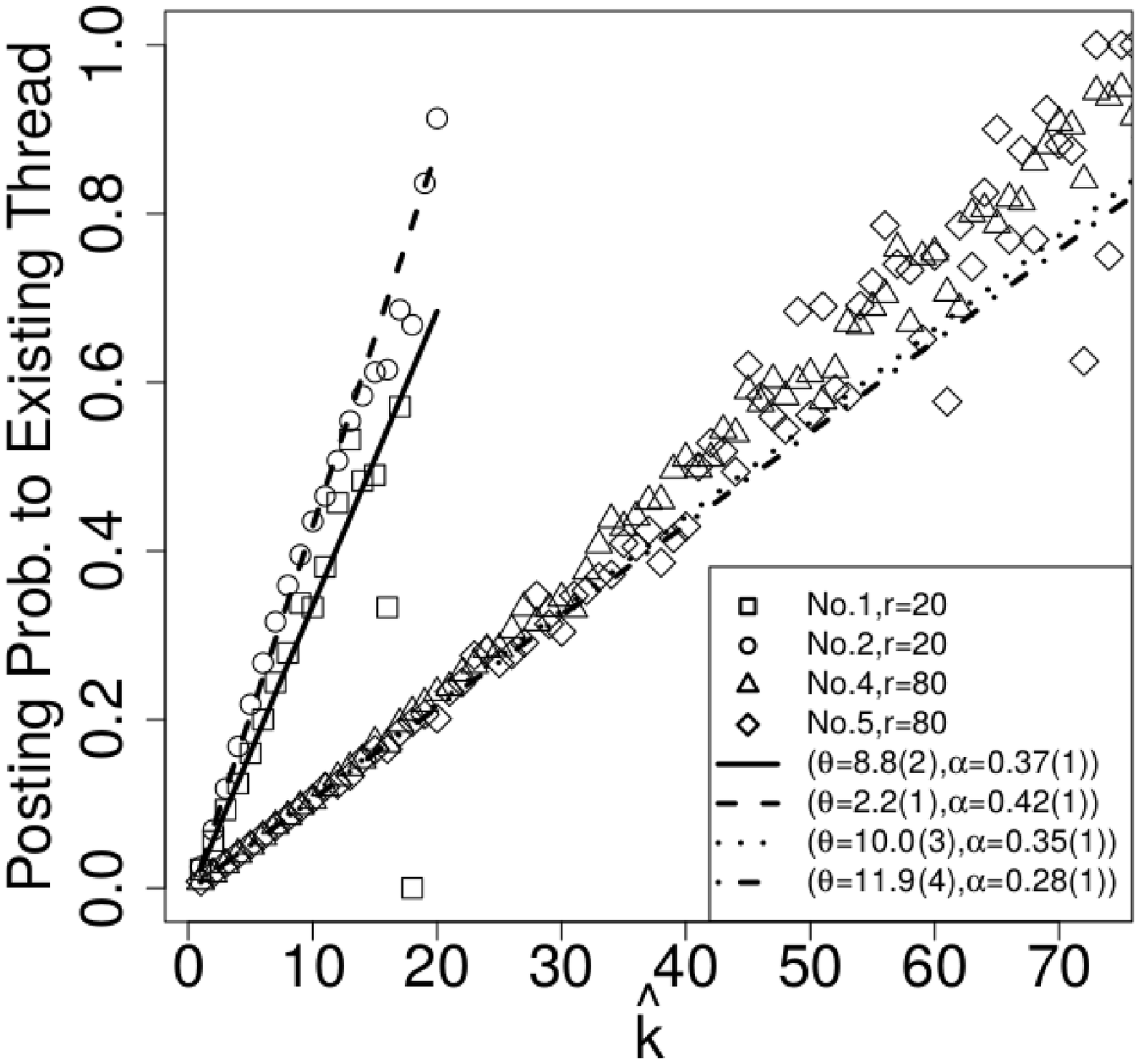} \\
\includegraphics[width=5.5cm]{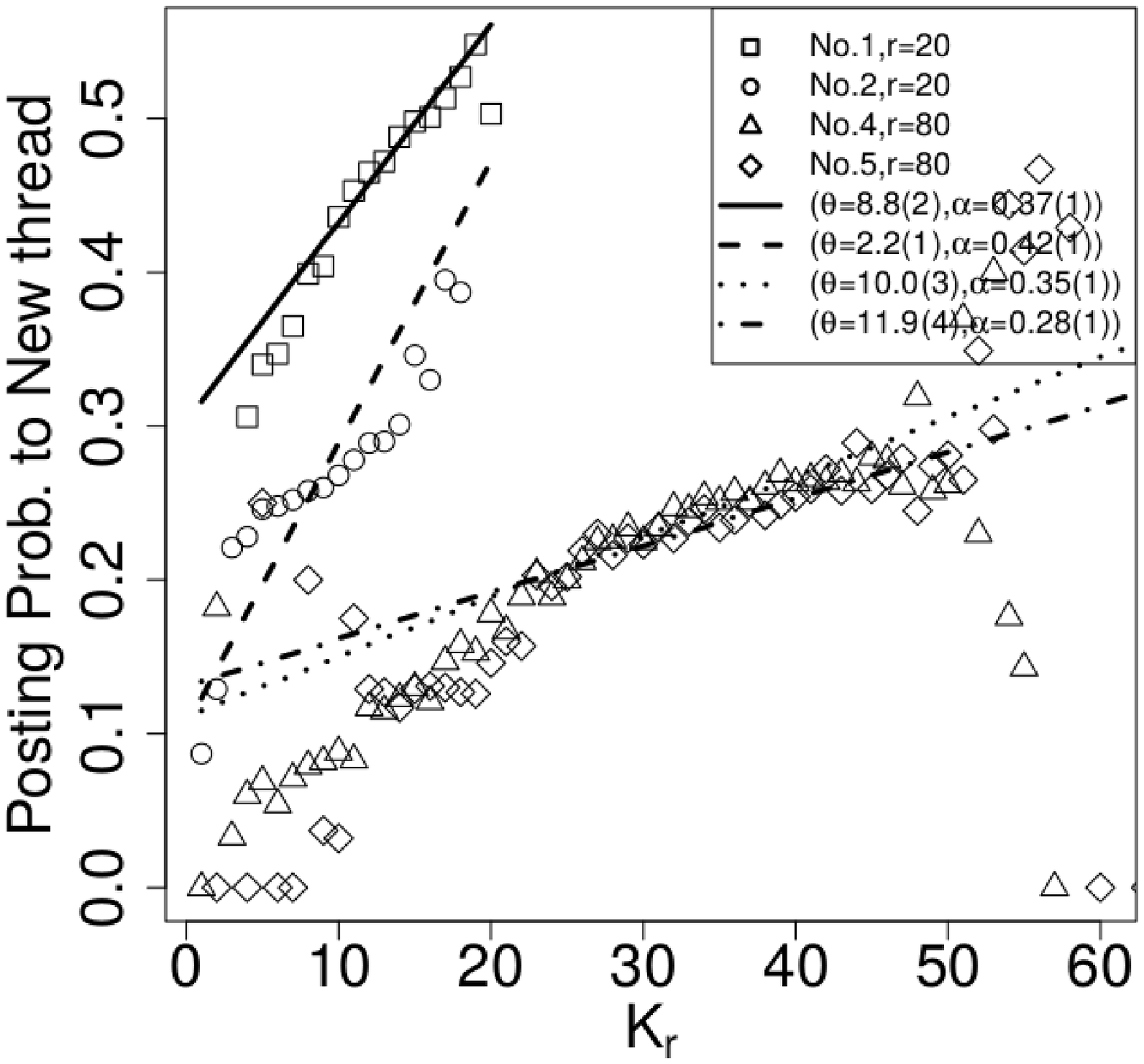} 
\end{tabular}
\caption{Plots of $P_{Existing}(\hat{k})$ vs. $\hat{k}$ and $P_{New}(K_r)$ vs. $K_r$.
  The symbols denote the estimated results using 
  $\hat{P}_{Existing}(\hat{k})$ in (\ref{eq:Fit1}) and $\hat{P}_{K_r}(K_r)$ in (\ref{eq:Fit2}).
  The lines denote the plots of
  (\ref{eq:2ch_1}) and (\ref{eq:2ch_2})
  with fitted parameters for $\theta$ and $\alpha$ in Table \ref{tab:fit}. 
}
\label{fig:Pr}
\end{figure}

To verify the probabilistic rules, we
directly estimate $P_{Exissting}(\hat{k})$ and $P_{New}(K_{r})$. We calculate the number of threads
with post times $\hat{k}$ for the past $r$ posts and denote it as $N(\hat{k})$.
In addition, we count the number of times a post is made on an
exiting thread with posts $\hat{k}$ and denote it as $N_{post}(\hat{k})$.
The estimator for $P_{Existing}(\hat{k})$ is
\begin{equation}
\hat{P}_{Existing}(\hat{k})=\frac{N_{Post}(\hat{k})}{N(\hat{k})} \label{eq:Fit1}.
\end{equation}
Likewise, we count the number of threads $K_r$ 
and the number
of times a post is made on a new thread when the number of threads is $K_r$.
We denote them as $N_{K_r}(K_r)$ and $N_{New}(K_r)$, respectively.
The estimator for $P_{New}(K_r)$ is then denoted as
\begin{equation}
\hat{P}_{New}(K_r)=\frac{N_{New}(K_r)}{N_{K_r}(K_r)} \label{eq:Fit2}.
\end{equation}

Fig. \ref{fig:Pr} presents the estimates for 
$\hat{P}_{Existing}(\hat{k})$ and $\hat{P}_{new}(K_r)$.
We also plot $P_{Existing}(\hat{k})$ and $P_{New}(K_r)$
in (\ref{eq:2ch_1}) and (\ref{eq:2ch_2}) with fitted values
for $\theta$ and $\alpha$ in Table \ref{tab:fit}.
The estimated results for the maximum likelihood fit well with the
 results from the estimators $\hat{P}_{Existing}(\hat{k})$
and $\hat{P}_{New}(K_r)$.
The parameters fall in zone II, which was introduced in the previous section.
In this zone, it is difficult for a leader to appear.

\subsection{Distribution of $K_r$ and $\hat{k}$}

\begin{figure}[htbp] 
\begin{tabular}{c}
\includegraphics[width=5.5cm]{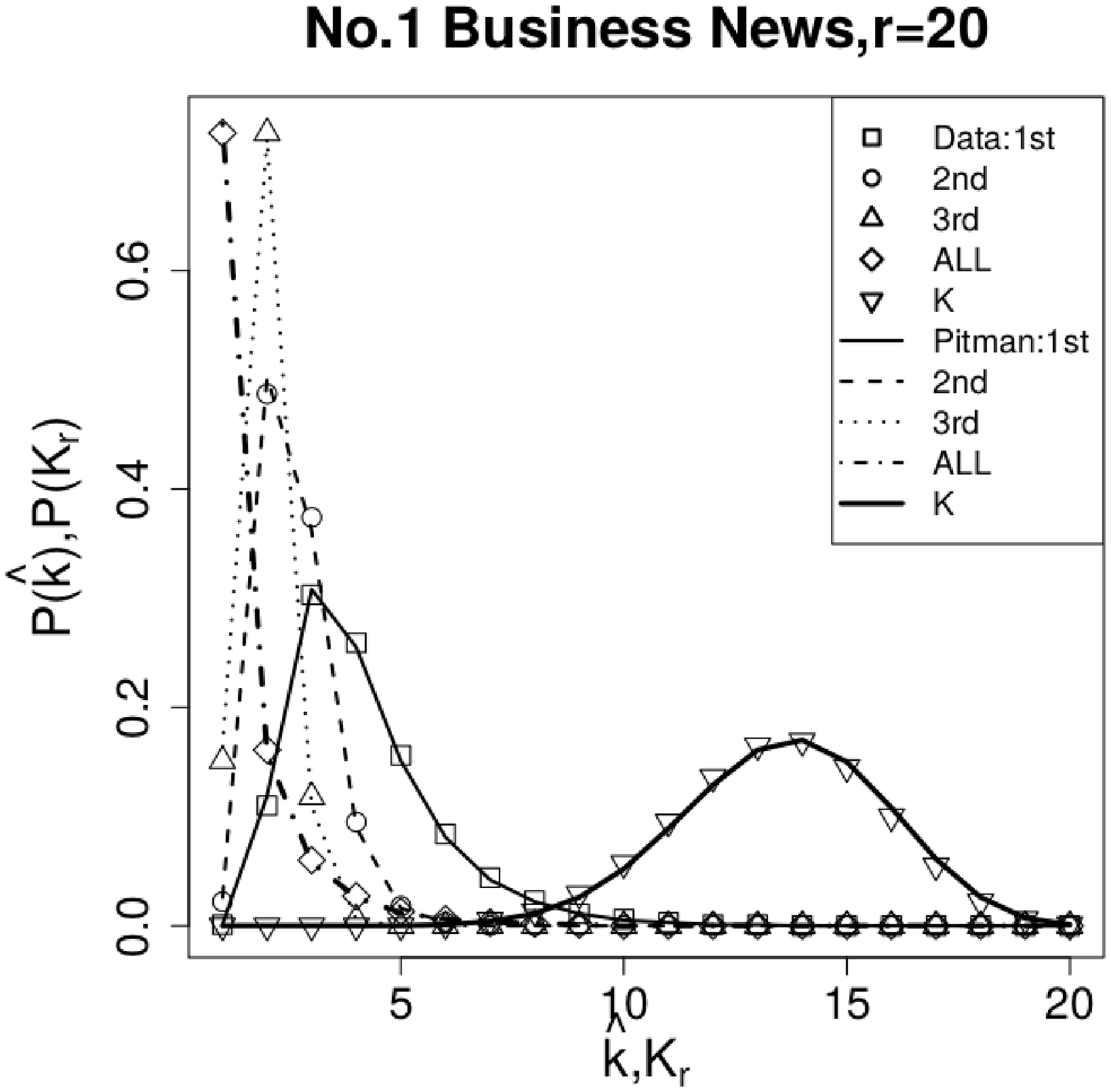} \\
\includegraphics[width=5.5cm]{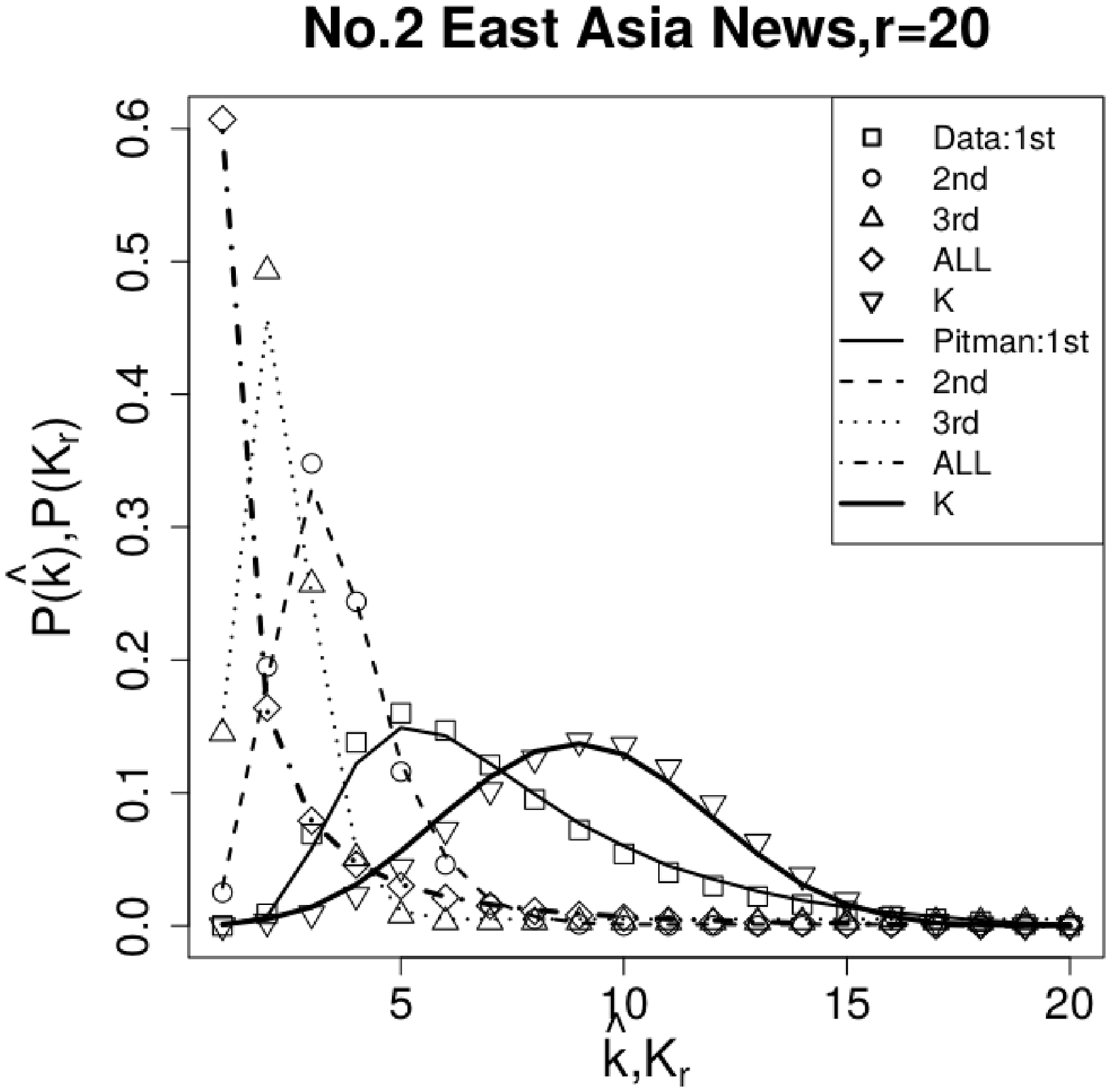} \\
\includegraphics[width=5.5cm]{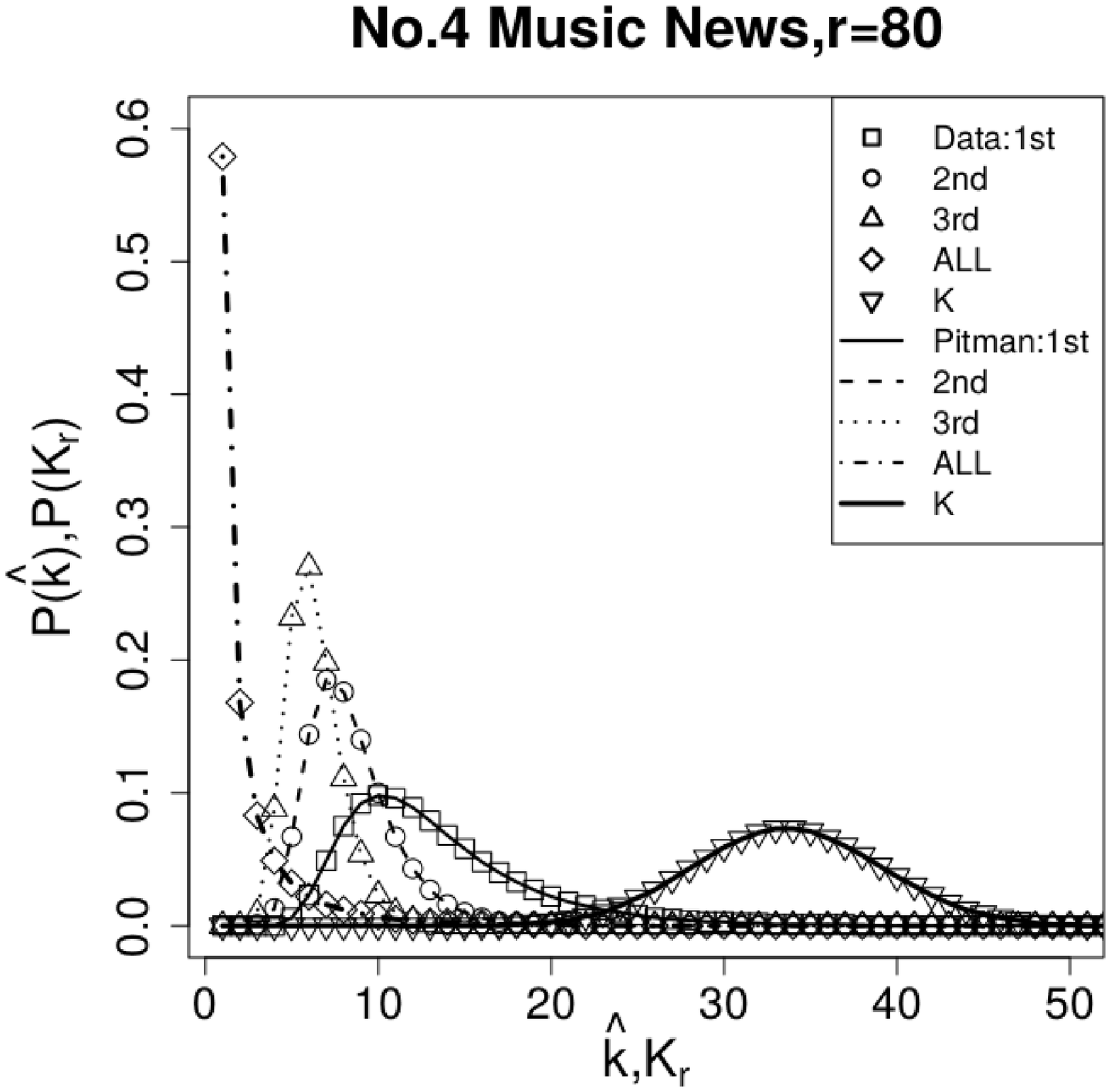} \\
\includegraphics[width=5.5cm]{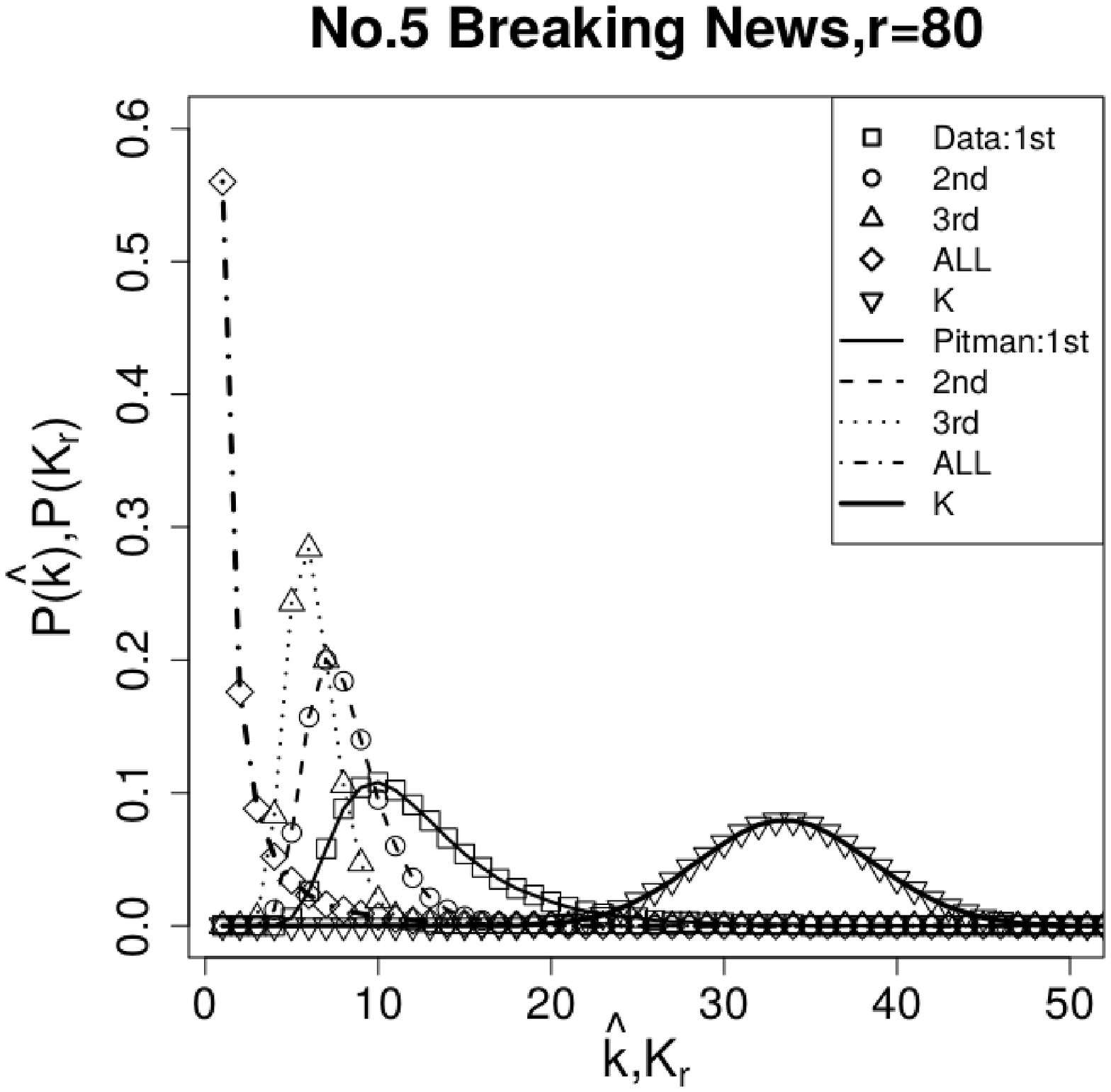} 
\end{tabular}
\caption{
  Plots of the distribution of post times $\hat{k}$ and number of threads $K_r$
  In the past $r$ posts for data with symbols and for (\ref{pit2}) and (\ref{pk}).
  We sort $\hat{k}$ in descending order and select the largest three values
  $\hat{k}_{1}\ge \hat{k}_{2}\ge \hat{k}_{3}$.
  The symbols $\Box,\circ$, and $\triangle$ denote
  the empirical distributions of $\hat{k}_{1},\hat{k}_{2}$, and $\hat{k}_{3}$,
  respectively.
  $\Diamond$ and $\bigtriangledown$ show the distribution
  of all $\hat{k}$ and $K_{r}$.
  The lines indicate the plots of $P_{r}(K_r)$ in (\ref{pk}) and
  the distributions of the ordered $\hat{k}_{j}$ and $\hat{k}$, which are calculated
  using the Pitman sampling formula in eq.(\ref{pit2}).
  We adopt $r=20 (80)$ for
  the No. 1 and No. 2 (No. 4 and No. 5) boards.
  The parameters for (\ref{pit2}) and (\ref{pk}) are presented in
  Table \ref{tab:fit}. We adopt the second set for $r=20$ and the first set
  for $r=80$.}
\label{fig:EPSF}
\end{figure}

We compare (\ref{pit2}) and the probability mass function 
for $K_r$ using the fitted parameters in Table \ref{tab:fit}
and those of an empirical distribution.
The probability mass function $P_{r}(K_r)$
for the number of candidates $K_r$ with reference $r$
is given as
\begin{eqnarray}
P_{r}(K_r)&=&\frac{\theta^{[r:\alpha]}}{\theta^{[r]}}c(r,K_r,\alpha)\alpha^{-K_r},
\label{pk} 
\end{eqnarray}
where $c(r,K_r,\alpha)$ 
is the generalized Stirling number or the C-numbers \cite{Pitman:2006}.
As for the probability mass function for the post times $\hat{k}$,
we calculate the number of posts $\hat{k}_{1st}\ge \hat{k}_{2nd}\ge \hat{k}_{3rd}$
for the most popular three threads in the past $r$ posts in addition to all
post times $\hat{k}$ for all threads.
We plot the results in Fig. \ref{fig:EPSF}.

As we can see, the fitting results are good. The distributions of 
$\hat{k}_{1st}$, $\hat{k}_{2nd}$,
 $\hat{k}_{3rd}$, $\hat{k}$, and $K_r$
are well described by the Pitman sampling formula (\ref{pit2}) and $P_{r}(K_r)$
with fitted parameters for $\theta,\alpha$ in Table \ref{tab:fit}.
We present the results for the statistical test using Kolmogorov--Smirnov (KS) statistics
in Appendix \ref{B}.

\subsection{Distribution of total votes $c_{n}(T)$}
\begin{figure}[tbh] 
\begin{tabular}{c}
\includegraphics[width=5.5cm]{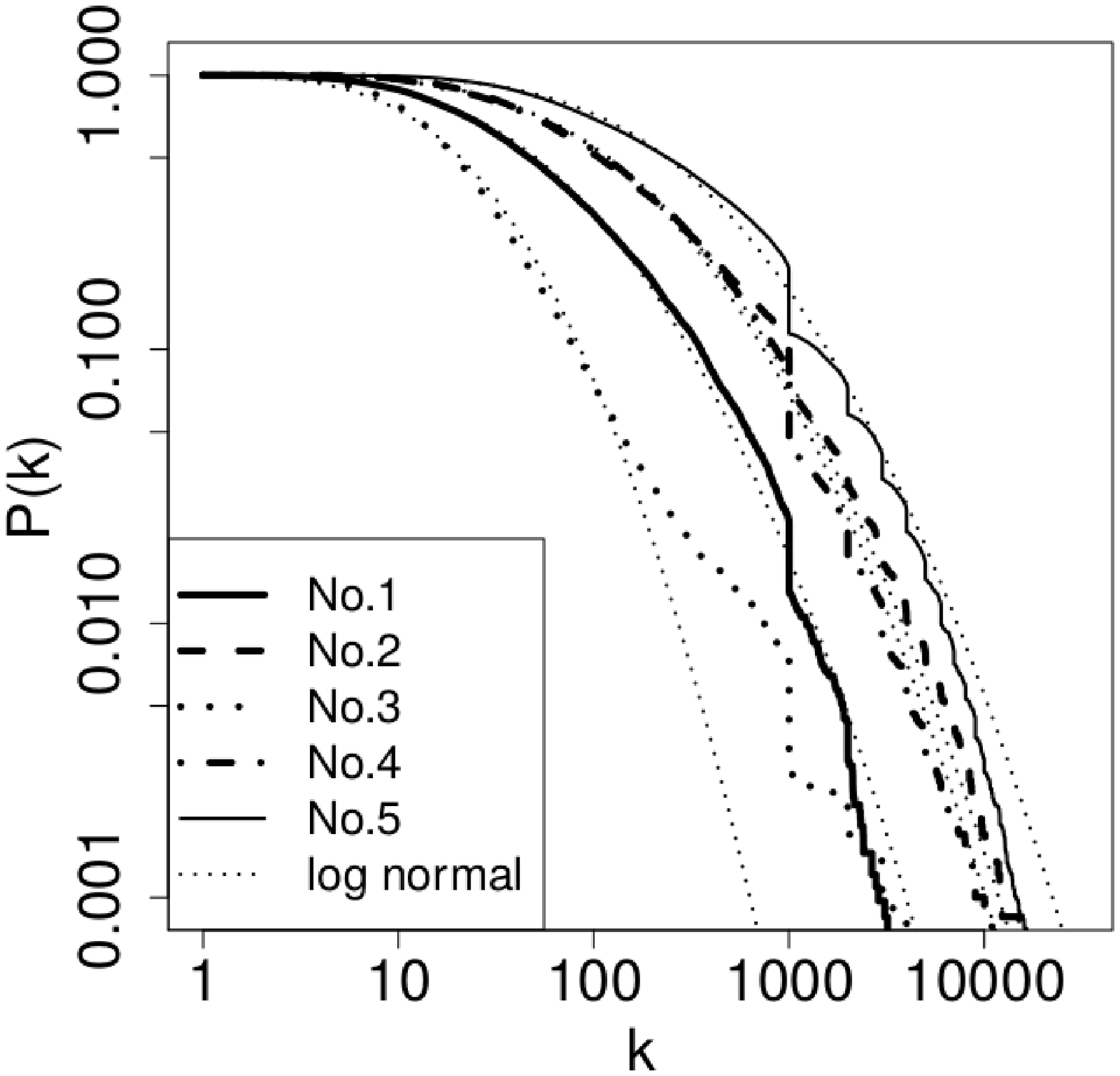} \\
\includegraphics[width=5.5cm]{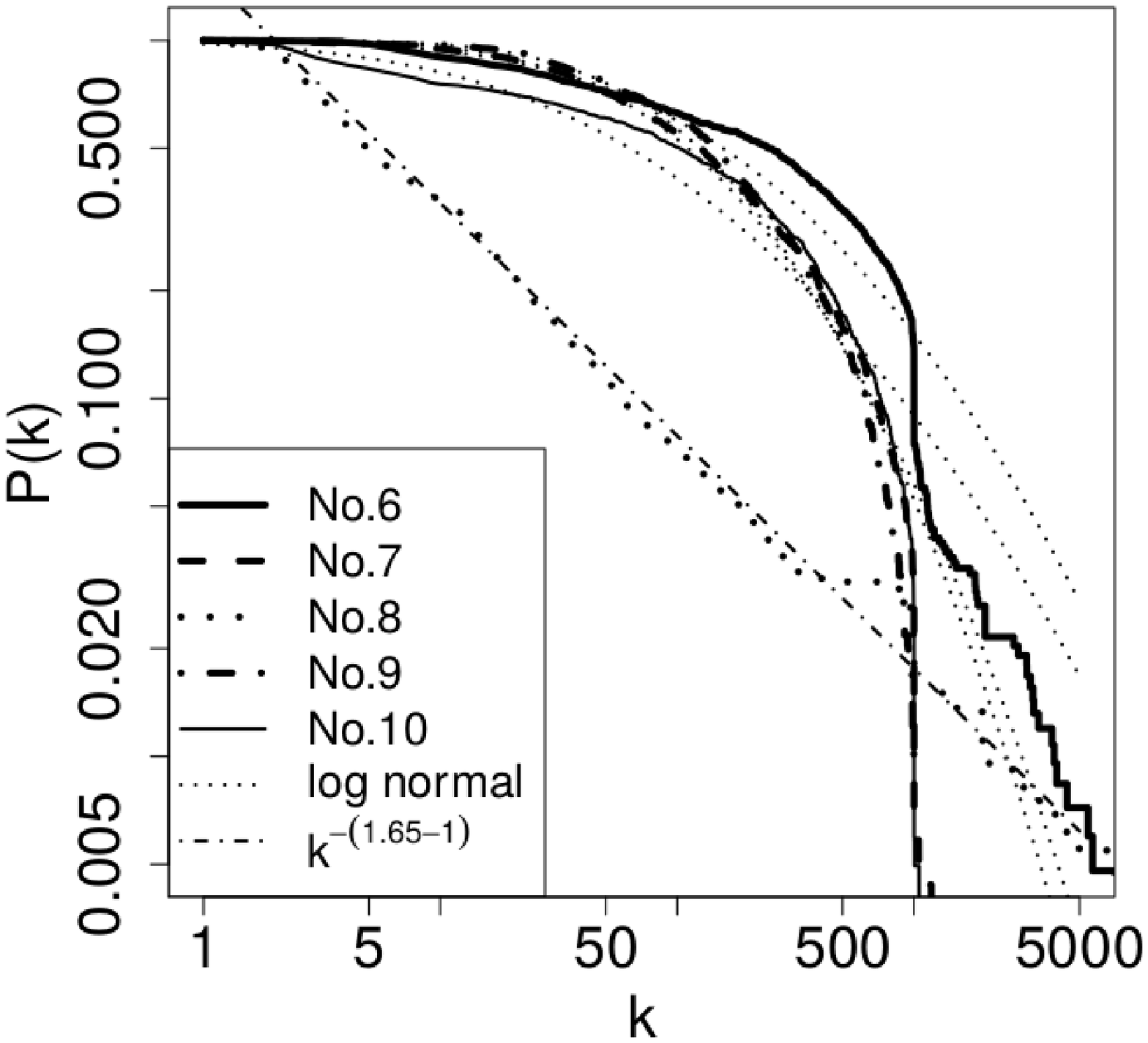} 
\end{tabular}
\caption{Plot of $P(k)\equiv \mbox{P}(c_{n}(T)\ge k)$ vs. $k$.
  The left (right) panel presents the results for the first (remaining)
  five boards. The means and standard deviations of $c_{j}$
  for the first five boards are $(3.93,1.41),(4.79,1.48),(3.03,1.10),(4.76,1.45)$, and $(5.47,1.47)$.
}
\label{fig:total}
\end{figure}

In this subsection, we discuss the distribution of total votes $c_n(T)$
for thread $n$, where $c_{n}(t)=\sum_{t'=1}^{t}\delta_{n(t'),n}$.
Fig. \ref{fig:total} shows the semi-logarithmic plot
of the cumulative distribution $P(k)\equiv \mbox{P}(c_{n}(T)\ge k)$ vs. $k$.
The left (right) panel depicts the results for the first (remaining) five boards.
The dotted line denotes the cumulative distribution of the log-normal distribution with the
same mean and variance. As is clearly shown, the results show good fits.
In the previous subsection, we confirm that the distribution of posts 
on the four boards obey the equilibrium Pitman sampling formula.
The difference between the equilibrium and
non-equilibrium suggests that the posting process for a large $r$
is not described by the voting process.

\begin{figure}[tbh] 
\begin{tabular}{c}
\includegraphics[width=5.5cm]{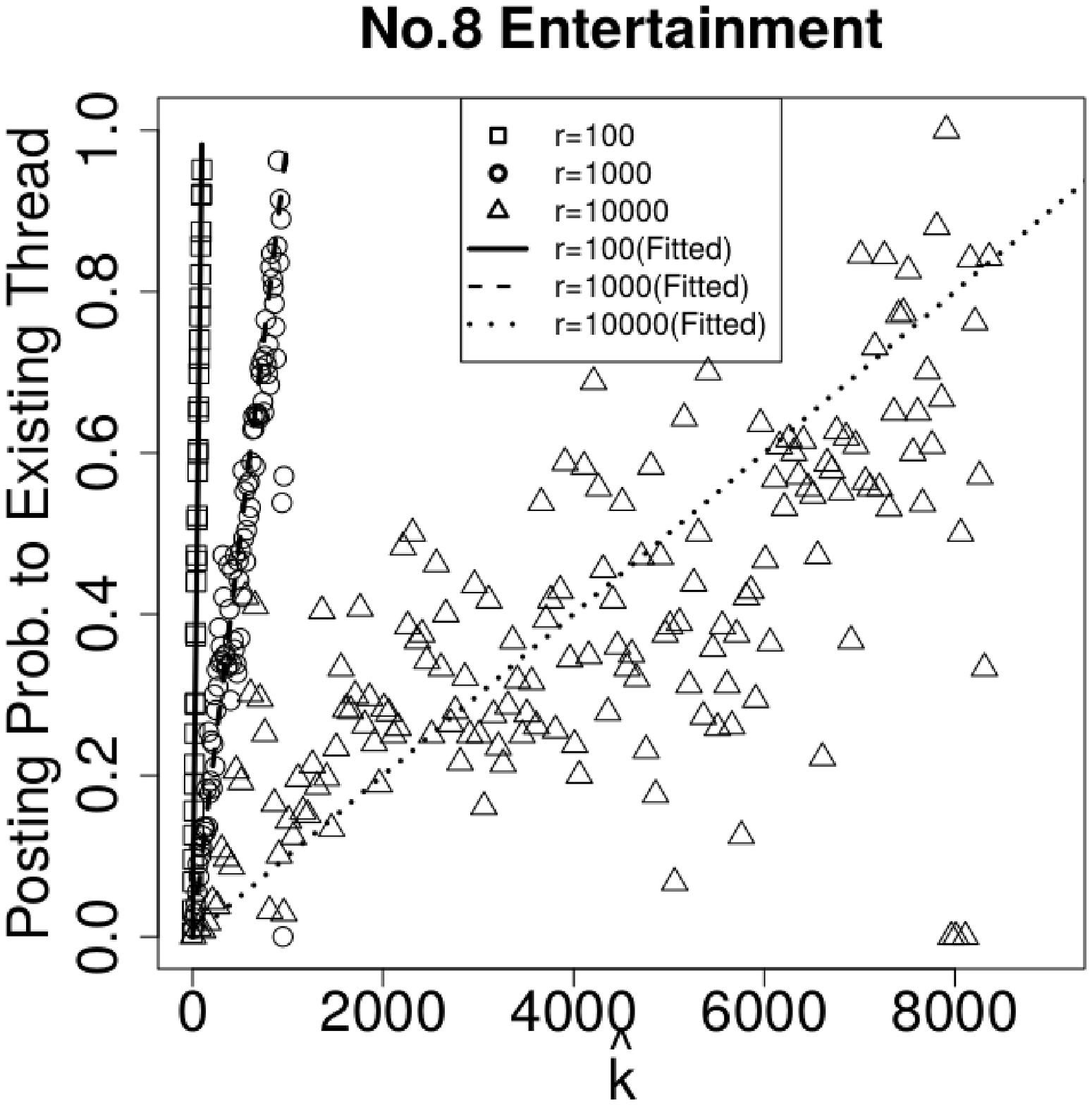} \\
\includegraphics[width=5.5cm]{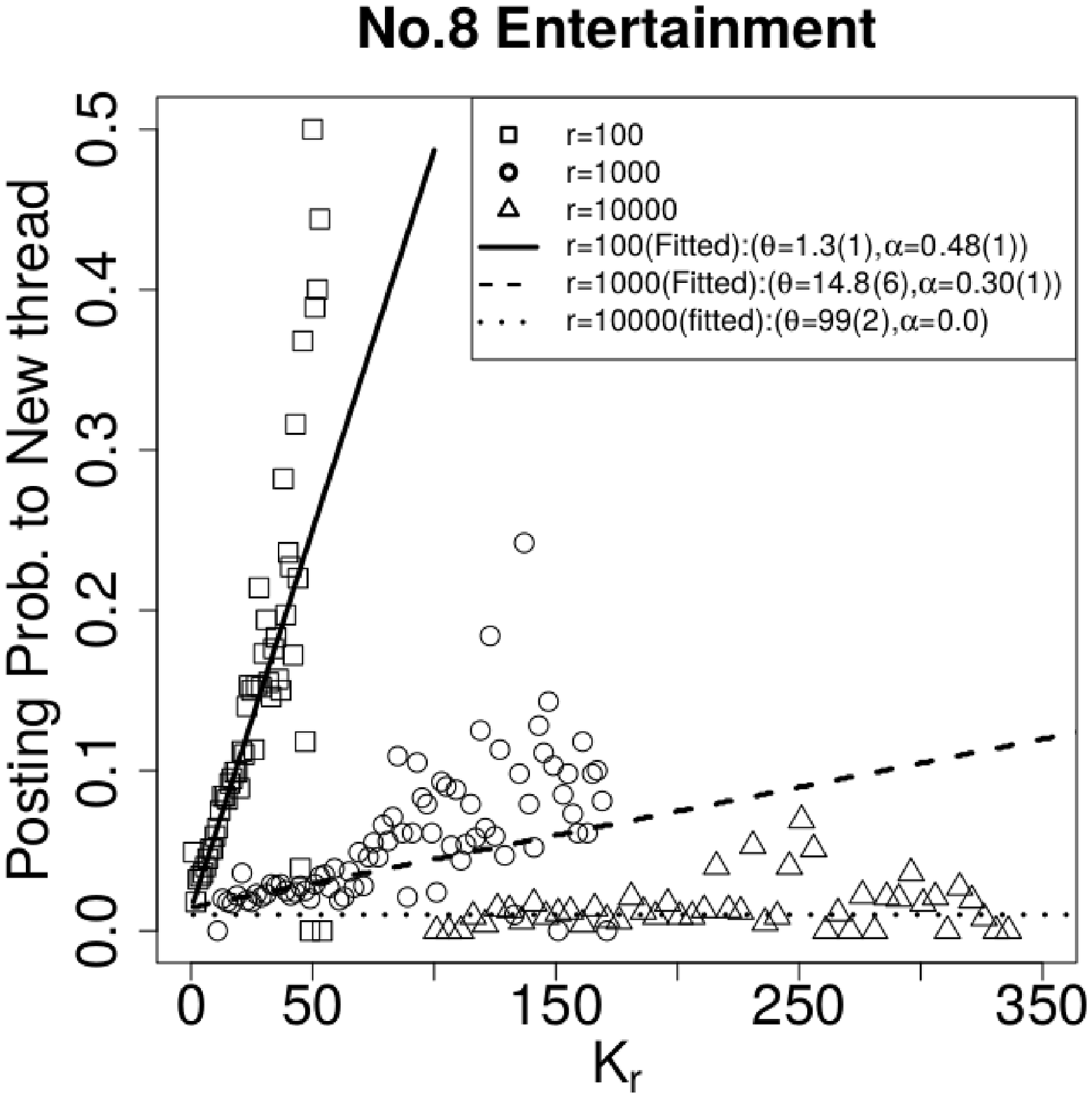} 
\end{tabular}
\caption{Plot of $P_{Existing}(\hat{k})$ vs. $P_{New} (K_r)$ for the No. 8 board. 
The symbols show the estimated results using 
  $\hat{P}_{Existing}(\hat{k})$ in (\ref{eq:Fit1}) and $\hat{P}_{K_r}(K_r)$ in (\ref{eq:Fit2}).
  The lines denote the plots of
  (\ref{eq:2ch_1}) and (\ref{eq:2ch_2}).}
\label{fig:No.8}
\end{figure}

In the remaining boards, we confirm that the distribution of the number of
posts on the No. 8 board follow
the power-law distribution, and the power-law index is 1.65, $P(k)\sim k^{-(1.65-1)}$.
We have seen that the board has a long memory in Fig.\ref{fig:Cor}.
Furthermore, we can confirm that the probability of a post is proportional to the 
number of posts and that of a new thread
is proportional to the number of threads for a large $r$ in Fig.\ref{fig:No.8}.
As $r$ increases, the latter dependence disappears, and the process is described
by the Yule process \cite{Yule:1925}. As the power-law exponent
is less than two, the fitness model for evolving networks might be
a better candidate to describe the
posting process \cite{Bianconi:2001,Hisakado:2016}.

\section{\label{sec:con} Concluding Remarks}
In this study, we discuss choice behavior using a voting model
comprising voters and candidates.
Voters vote for a candidate with a probability 
that is proportional to the ratio of previous votes,
which is visible to the voters.
In addition, voters can obtain information from a finite
number $r$ of the most recent previous voters.

In the large $t$ limit, the system is equilibrated, and
the partition of $r$ votes follows the Pitman sampling formula.
Kirman's ant colony model is a special case that corresponds to
the number of states $K=2$.	
The equilibrium probability distribution and the non-equilibrium
probability distribution for $t=r$ are the same.
 We propose using this voting model for the posting process of a BBS, 2ch.net,
 where users can select one of many threads to make a post. We
explore how this choice depends on the last 
$r$ posts and the distribution of the last $r$ posts across boards.
We conclude that the posting data in the news category is
described by the voting model.
The equilibrium time or time horizon $s_{H}$ is about 1.5-4 minutes.
Up to this time horizon, the probability
of posting on a thread is proportional to the ratio of
posts on the thread.

When the number of candidates $K$ is fixed at
$\theta=-K\alpha$ for $\alpha<0$, we show that the Dirichlet multinomial distribution 
 reduces to the Pitman sampling formula in Appendix \ref{A}.
 In Appendix \ref{C}, we show an application to
 parliament election data for Japan.
 We ignore the inhomogeneities among the candidates in the elections.
 The model has only one parameter $\theta$.
 We estimate the correlation strength between votes as $1/(\theta+1)$ and
 show that the correlation between votes becomes stronger 
 after the introduction of the small constituency system in 1993.

 Recently, a wide variety of social systems, including
 election votes, citations of scientific papers,
 rating dynamics on E-commerce, and social tagging systems, have been extensively studied
 \cite{Gracia:2014,Wan:2013,Wan:2014,Cattuto:2009,Hashimoto:2016} using simple
 probabilistic models. 
 We hope our study provides a new perspective from an equilibrium viewpoint in a
 non-equilibrium system.

\begin{acknowledgment}
  H.M. performed the theoretical analysis. S.M. conducted the analysis of the
  post data for 2ch.net, and F.S. did so for the election data
  in Appendix \ref{C}.
  All authors contributed to the analysis and interpretation of the results and
  the writing of the manuscript. This work is supported by JPSJ KAKENHI[Grant No. 17K00347]. 
\end{acknowledgment}

\bibliographystyle{jpsj} 
\bibliography{68411}

\appendix

\section{\label{A} Fixed number of candidates case}

We model the voting of $K$ candidates, $C_1\cdots C_K$.
At time $t$, candidate $C_j$ has $c_j(t)$ votes.
In this appendix, we consider the case in which the number of
candidates $K$ is fixed, that is, no new entry is allowed. 
In each time step, one voter votes for one candidate;
 the voting is sequential.
Hence, at time $t$, the $t$th voter votes, after which the total number of votes is $t$.
Voters are allowed to see  
$r$ previous votes for each candidate and, thus, are aware of public perception.
$r$ is a constant number.
We consider the case in which all voters vote
for the candidate with a probability proportional to the previous votes
ratio, which is visible to the voters.

The transition is 
\begin{equation}
c_j(t)=k \rightarrow k+1:
 P_{j,k,t:l,t-r}=\frac{\frac{q_j(1-\rho)}{\rho}+ (k-l)}{\frac{1-\rho}{\rho}+ r}=\frac{\beta_j+(k-l)}{\theta+ r},
\label{pda}
\end{equation}
where $c_j (t-r)=l$, $\rho$ is the correlation coefficient, and
$q_j$ is the initial constant of the $j$th candidate \cite{Hisakado:2006}.
$\rho$ is the correlation of the beta binomial model.
The constraint $\sum_{j=1}^{K}q_j=1$ exists.
 We define $\theta=(1-\rho)/\rho$ and $\beta_j=q_j (1-\rho)/\rho$.
$P_{j,k,t:l,t-r}$ denotes the probabilities of the process.
The voting ratio for $C_j$ at $t-r$ is $c_j (t-r)=l$.
We consider the case $\beta_j\geq 0$ from $P_{j,k,t:l,t-r}>0$ and 
the constraint $\sum_j \beta_j=\theta$.
 When $\beta_j=\beta$, the constraint becomes $\beta K=\theta$.

We consider the hopping rate among $(r+1)$ states $\hat{k}_j=k-l$, $\hat{k}_j=0,1,\cdots, r$. In each step of $t$, the vote at time $(t-r)$ is deleted, and a new vote is obtained.    
$\hat{k}$ is the number of votes candidate $C_j$ obtained in the latest $r$ votes.
In case $K=2$, the model becomes Kirman's ant colony model \cite{Kirman:1993}.
The dynamic evolution of the process is given by
\begin{eqnarray}
\hat{k}_j&\rightarrow& \hat{k}_j+1:
 P_{\hat{k}_j,\hat{k}_j+1,t}=\frac{r-\hat{k}_j}{r}\frac{\beta_j+ \hat{k}_j}{\theta+ r-1},
\nonumber \\
\hat{k}_j &\rightarrow& \hat{k}_j-1:
P_{\hat{k}_j,\hat{k_j}-1,t}=\frac{\hat{k}_j}{r}\frac{(\theta-\beta_j)+ (r-1-\hat{k}_j)}{\theta+ r-1},
\nonumber \\
\hat{k}_j &\rightarrow& \hat{k}_j:
P_{\hat{k}_j,\hat{k}_j,t}=1-P_{\hat{k},\hat{k}-1,t}-P_{\hat{k},\hat{k}+1,t}.
\nonumber
\end{eqnarray}
$P_{\hat{k}_j,\hat{k}_j\pm 1,t}$ are the probabilities of the process
and the products of exit votes and new entry votes.

We consider hopping from candidate $C_{i}$ to $C_{j}$.
\begin{eqnarray}
\hat{k}_i&\rightarrow& \hat{k}_i-1,\hat{k}_j \rightarrow \hat{k}_j+1:
 P_{\hat{k}_i\rightarrow\hat{k}_i-1,\hat{k}_j\rightarrow\hat{k}_j+1,t}=\frac{\hat{k}_i}{r}\frac{\beta_j+ \hat{k}_j}{\theta+ r-1},
\nonumber \\
\hat{k}_i&-&1\rightarrow \hat{k}_i,\hat{k}_j +1\rightarrow \hat{k}_j:
P_{\hat{k}_i-1\rightarrow\hat{k}_i,\hat{k}_j+1\rightarrow\hat{k}_j,t} \nonumber \\
&=&\frac{\hat{k}_j+1}{r}\frac{\beta_i+ \hat{k}_i-1}{\theta+ r-1}.
\nonumber
\end{eqnarray}
Here, we define $\mu_{r}(\hat{k},t)$ as a distribution function of
the state $\hat{k}$ at time $t$.
The number of all states is $(r+1)$. Given that the process is
reversible, we have
\begin{equation}
\frac{\mu_{r}(\hat{k}_i,\hat{k}_j, t)}{\mu_{r}(\hat{k}_i-1,\hat{k}_j+1, t)}
=\frac{\hat{k}_j+1}{\hat{k}_i}\frac{\beta_i+\hat{k}_i-1}{\beta_j+\hat{k}_j}.
\end{equation}
We can separate indexes $i$ and $j$ and obtain
\begin{eqnarray}
\frac{\mu_{r}^i(\hat{k}_i, t)}{\mu_{r}^i(\hat{k}_i-1, t)}
&=&\frac{\beta_i+\hat{k}_i-1}{\hat{k}_i}c
\nonumber \\
\frac{\mu_{r}^j(\hat{k}_j+1, t)}{\mu_{r}^j(\hat{k}_j, t)}
&=&\frac{\beta_j+\hat{k}_j}{\hat{k}_j+1}c,
\label{se}
\end{eqnarray}
where $c$ is a constant.
Using (\ref{se}) sequentially, in the limit $t\rightarrow \infty$,
we can obtain the equilibrium distribution, which can be written as
\begin{equation}
\mu_r(\bm{a}
,\infty)=
\left(
\begin{array}{r}
\theta+r-1\\
r
\end{array}
\right)^{-1}
\prod_{j=1}^{K}
\left(
\begin{array}{r}
\beta_j +\hat{k}_j-1\\
\hat{k}_j
\end{array}
\right),
\label{bin}
\end{equation}
where $\bm{a}=(\hat{k}_1,\hat{k}_2,\cdots, \hat{k}_K)$.
This distribution is written as 
\begin{equation}
\mu_r(\bm{a},\infty)=\frac{r !}{\theta^{[r]}}\prod_{i=1}^{K} \frac{\beta_i^{[\hat{k}_i]}}{\hat{k}_i !},
\label{D}
\end{equation}
where $x^{[n]}=x(x+1)\cdots (x+n-1)$.
This is the Dirichlet multinomial distribution.

Here, we set $\beta_j=\beta$.
The relation $\beta=-\alpha$ exists,
where $\alpha$ is the parameter used in the main text.
We write (\ref{bin}) as
\begin{equation}
\mu_r(\bm{\hat{a}},\infty)=
\left(
\begin{array}{r}
\theta+r-1 \\
r
\end{array}
\right)^{-1}
\prod_{j=1}^{r}
\left(
\begin{array}{r}
\beta+j-1 \\
j
\end{array}
\right)^{a_{j }},
\end{equation}
where $a_{j }$ is the number of candidates for whom $j$ voters voted and 
$\bm{\hat{a}}=(a_1, \cdots, a_{r})$.
Hence, the relations $\sum_{i=1}^{r}a_i=K_r<K$ and $\sum_{i=1}^{r}i a_i=r$ exist.
Here, we define $K_r$ as the number of candidates who have more than one vote.
$(K-K_r)$ candidates have no vote.
  
We consider the partitions of integer $K_r$. 
To normalize, we add the term of combination: $K!/a_1 !\cdots a_r !(K-K_r )!$.
We obtain
\begin{eqnarray}
&&\mu_r(\bm{\hat{a}},\infty) \nonumber \\
&=&
\frac{K!}{a_1 !\cdots a_r !(K-K_r )!}
\left(
\begin{array}{r}
\theta+r-1 \\
r
\end{array}
\right)^{-1}
\prod_{j=1}^{r}
\left(
\begin{array}{r}
\beta+j-1 \\
j
\end{array}
\right)^{a_{j }}
\nonumber \\
&=&
\frac{r! \theta^{[K_r:-\beta]}}{\theta^{[r]}}
\prod_{j=1}^{r}(\frac{(1+\beta)^{[j-1]}}{j!})^{a_j}\frac{1}{a_j!},
\label{pit}
\end{eqnarray}
where
$x^{[n:-\beta]}=x(x-\beta)\cdots (x-(n-1)\beta)$.
We use the relation $\theta=K\beta$.
(\ref{pit}) is simply the Pitman sampling formula \cite{Pitman:2006}.

In the limit $\beta\rightarrow 0$ and $K\rightarrow \infty$, subject to a fixed $\theta=\beta K$,
we can obtain the Ewens sampling formula.
In this case, the sum of the probabilities that a candidate who has zero votes can obtain
one vote is $\theta/(\theta+r)$. This case is the same as $\alpha=0$ in Section 2.

\section{\label{C} Data analysis of election data}
In this appendix, we study the distribution of vote shares in
the elections for Japan's House of Representatives.
Previously, we proposed a mechanical model that is based on a
voting model and certain assumptions about the inhomogeneities
of candidates \cite{Sano:2017}.
We describe the vote shares of candidates of a political party as the mixture
of the votes of fixed supporters of the political party in the region
and the votes of floating voters, which obeys the Dirichlet distribution.
The former group becomes a source of the inhomogeneity of the system.
Here, we neglect the inhomogeneities and treat all candidates equally.

\subsection{Election data}
We study data from the 28th general election
in 1958 to the 47th general election in 2014 and find a change in the election system.
A middle constituency system was adopted prior to the 40th election such that
the number of winners in each district is generally between three and five. In the case of the correction of congress seats,
there are districts with 2--6 congress seats, which are rare cases. Following the
40th election in 1993, a small constituency system was
installed, and only one person was elected from each district.
In the analysis of election data, we used a dataset \cite{Mizusaki:2015}
that records the election results in several small regions
in each electoral district. We separate the election data before and after the
introduction of the small constituency system. Next, we classify data
on the basis of the number of congress seats $W$ and 
that of candidates $K$. Table \ref{tab:election} shows the
sample number case $(W,K)$.
Hereafter, we only study the case $(W,K)$ in which the
sample number is more than $10^{3}$.
For $W=1$, a small constituency system, we study $K=3,4,5$.
For $2\le W\le 5$, a middle constituency system, we examine
$K=4,5,6,7$ for $W=3$, $K=6,7,8,9$ for $W=4$ and
$K=6,7,8,9,10$ for $W=5$, respectively.

\begin{table}[htbp]
\caption{Number of samples in case $(K,W)$. 
  Election data are classified by the number of congressional seats
  $W$ and that of candidates $K$
  in the electoral districts. 
  The second column presents the results 
  for the elections under the small constituency system.
  The third, fourth, and fifth columns
  provide the results for $W=3,4,5$
  for the elections under the middle constituency system.}
\label{tab:election}
\begin{center}
\begin{tabular}{|l|c||c|c|c|} \hline
$K \backslash W $ & 1  & 3 & 4 & 5  \\ \hline
           \ \ \ \ \,2 & 814  &   NA & NA   & NA   \\  
           \ \ \ \ \,3 & 9,868 & NA   & NA   & NA   \\ 
           \ \ \ \ \,4 & 5,650 & 1,050 & NA   & NA   \\
           \ \ \ \ \,5 & 1,762 & 4,413 & 846  & NA   \\
           \ \ \ \ \,6 & 428  & 3,422 & 3,638 & 1,062 \\  
           \ \ \ \ \,7 & 115  & 1,652 & 4,104 & 4,613 \\
           \ \ \ \ \,8 & 18   & 562  & 2,850 & 5,358 \\
           \ \ \ \ \,9 &  6   & 227  & 1,112 & 4,511 \\
         \ \ \ \ \,10 &  NA   & 65   & 497  & 2,369 \\ 
         \ \ \ \ \,11 &  NA   & 81   & 148  & 796  \\ \hline
\end{tabular}
\end{center}
\end{table}

\subsection{Parameter estimation}
In the elections, the number of candidates $K$ is fixed and 
 the model parameters $\theta,\alpha$ should
satisfy $\alpha<0$ and $\theta=-K \alpha$.
We estimate the model parameter $\theta$
using the maximum likelihood principle.
As for $r$, we adopt $r=50,100$ and
the Dirichlet limit $r\to \infty$. 
For a finite $r$, we
transform the vote share $v_{i},i=1,\cdots,K$ into
votes $\hat{k}_{i},i=1,\cdots,K$ as $\hat{k}_{i}=\lfloor r\cdot v_{i}\rfloor$.
Here, $\lfloor x\rfloor$ is the floor function.
As for the treatment of fractions $r\cdot v_{i}-\hat{k}_{i}$,
we distribute the remaining votes $r-\sum_{i}\hat{k}_{i}$ to $\hat{k}_i$
with the largest fractions.

The results are presented in Table \ref{tab:election2}.
The estimated parameters $\theta$ for $(W,K)$
in the three cases are almost the same.
 The parameters are in zone III, which is introduced in Section III.
 In this region, it is difficult to be the stable leader.
 As the number of winners $W$ increases, the correlation decreases
 for the same number of candidates $K$.
 This finding means that the correlation strengthened following the
 introduction of the small constituency system in 1993.

\begin{table}[htbp]
\caption{$\theta$ for $(K,W)$,$r=50$}
\label{tab:election2}
\begin{center}
\begin{tabular}{|l|c||c|c|c|} \hline
$\theta(K,W)$& $W=1$     & 3         & 4         &   5           \\ \hline
$K=3$        & 3.2755(5) &           &           &               \\ 
4            & 3.9188(9) & 7.56(2)   &           &               \\
5            & 4.603(3)  & 7.150(4)  &           &               \\ 
6            &           & 6.146(3)  & 8.634(5)  & 14.81(7)      \\  
7            &           & 5.772(4)  & 7.575(3)  & 10.682(6)     \\
8            &           &           & 7.913(4)  & 10.366(4)     \\
9            &           &           & 7.401(8)  & 8.684(3)      \\
10           &           &           &           & 8.635(5)      \\  \hline
\end{tabular}
\end{center}
\begin{center}
\caption{$\theta$ for $(K,W)$,$r=100$}
\begin{tabular}{|l|c||c|c|c|} \hline
$\theta(K,W)$ & $W=1$        &   3       &    4     & 5         \\ \hline
  $K=3$       & 3.1341(4)    &           &          &           \\ 
  4           & 3.7544(8)    &  6.80(1)  &          &           \\ 
  5           & 4.414(2)     &  6.473(2) &          &           \\ 
  6           &              &  5.603(2) & 7.730(3) & 12.51(3)  \\  
  7           &              &  5.363(3) & 6.954(2) & 9.354(3)  \\
  8           &              &           & 7.169(3) & 9.078(2)  \\
  9           &              &           & 6.870(6) & 7.835(2)  \\
 10           &              &           &          & 7.781(3)  \\ \hline 
\end{tabular}
\end{center}
\begin{center}
\caption{$\theta$ for $(K,W)$ in the Dirichlet limit $(r\to \infty)$}
\begin{tabular}{|l|c||c|c|c|} \hline
$\theta(K,W)$ & $W=1$        &   3       &    4     & 5         \\ \hline
  $K=3$       & 3.0062(3)    &           &          &           \\ 
  4           & 3.568(1)    &  6.112(9) &          &           \\ 
  5           & 4.175(2)     &  5.786(1) &          &           \\ 
  6           &              &  5.060(1) & 6.906(2) & 10.38(2)  \\  
  7           &              &  5.000(1) & 6.289(1) & 8.017(2)  \\
  8           &              &           & 6.463(2) & 7.804(1)  \\
  9           &              &           & 6.376(3) & 7.110(1)  \\
 10           &              &           &          & 7.129(2)  \\ \hline 
\end{tabular}
\end{center}
\end{table}

\begin{figure*}[htbp]
\begin{tabular}{ccc}    
\includegraphics[width=5.5cm]{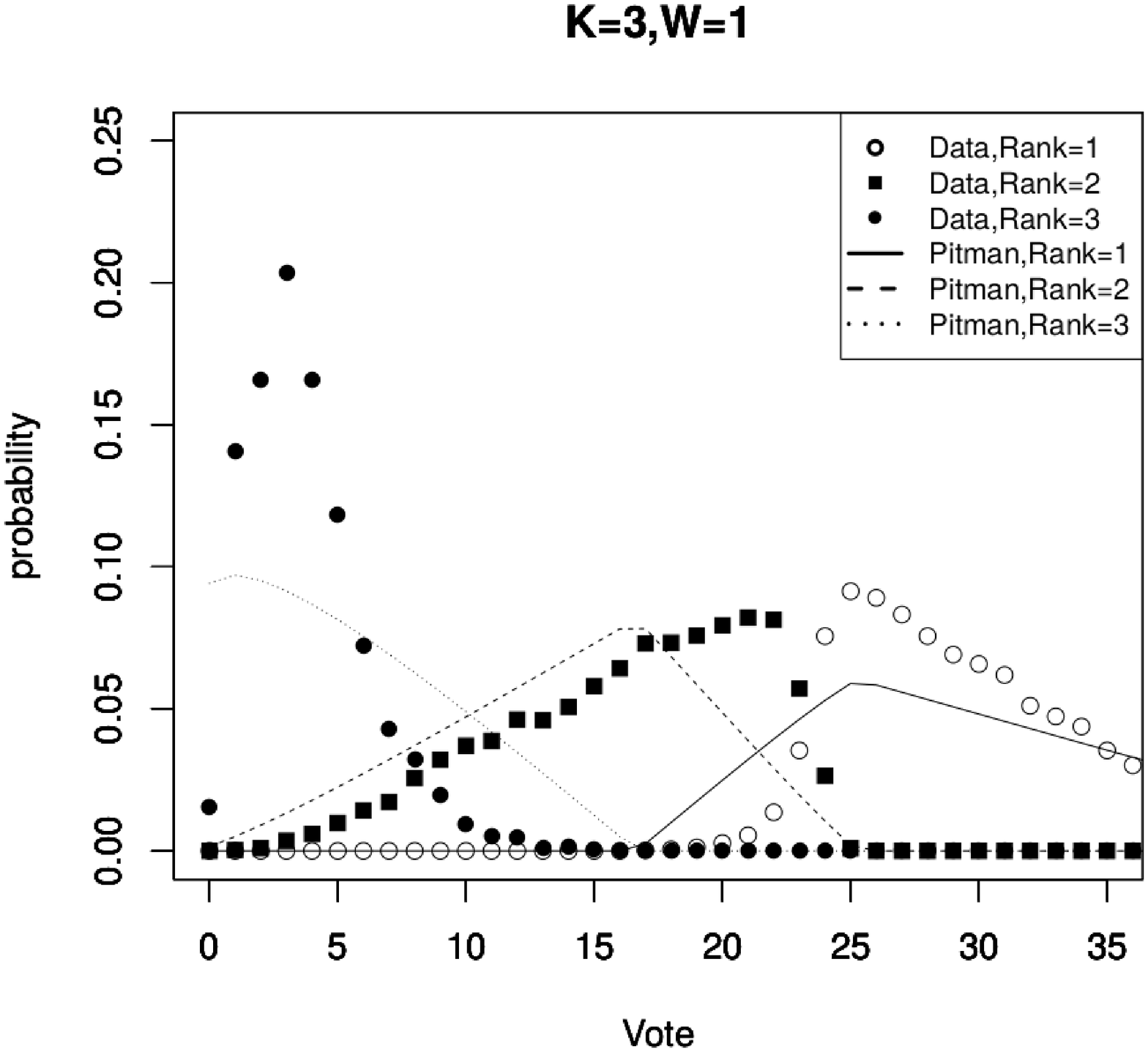} &
\includegraphics[width=5.5cm]{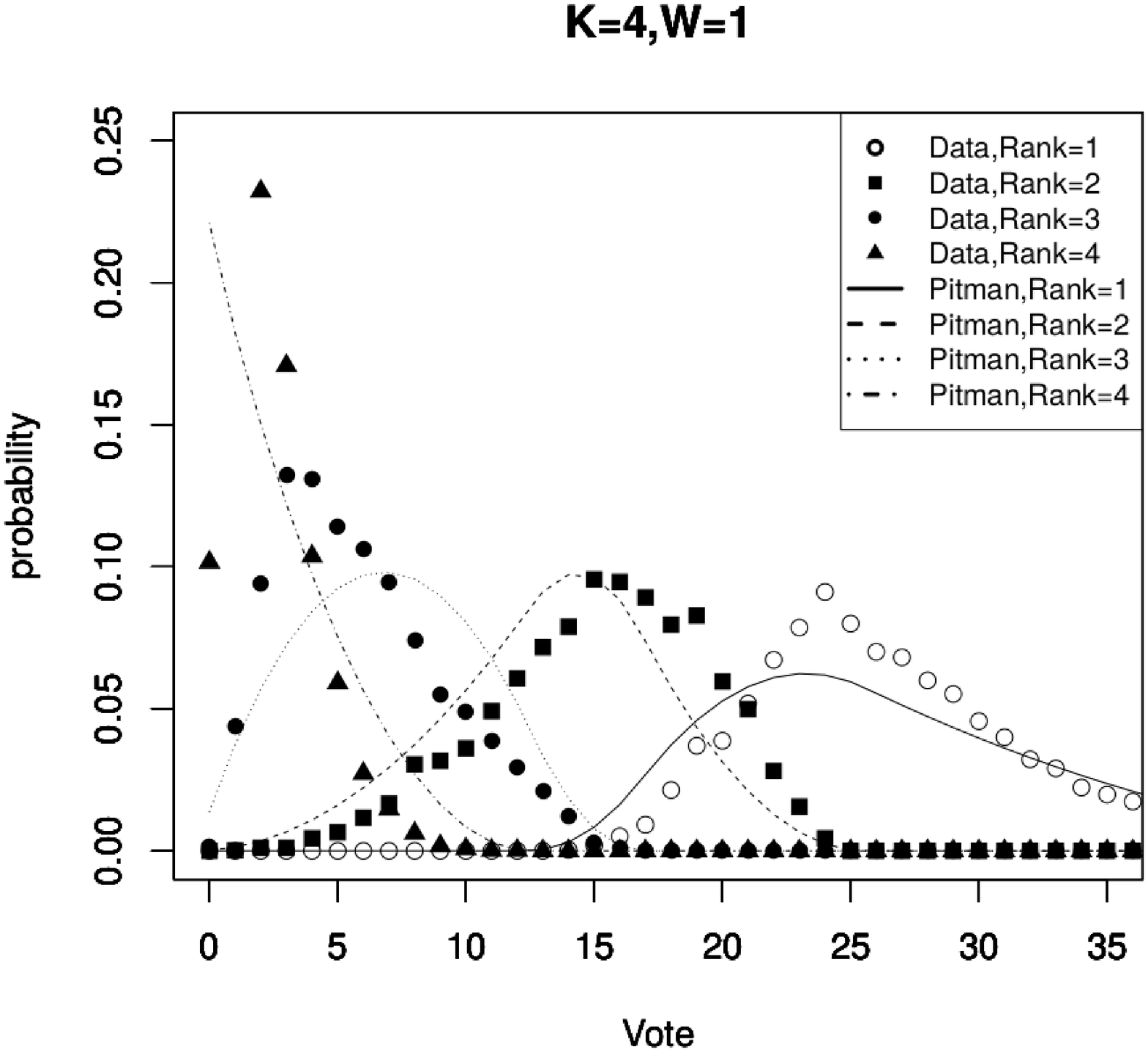} &
\includegraphics[width=5.5cm]{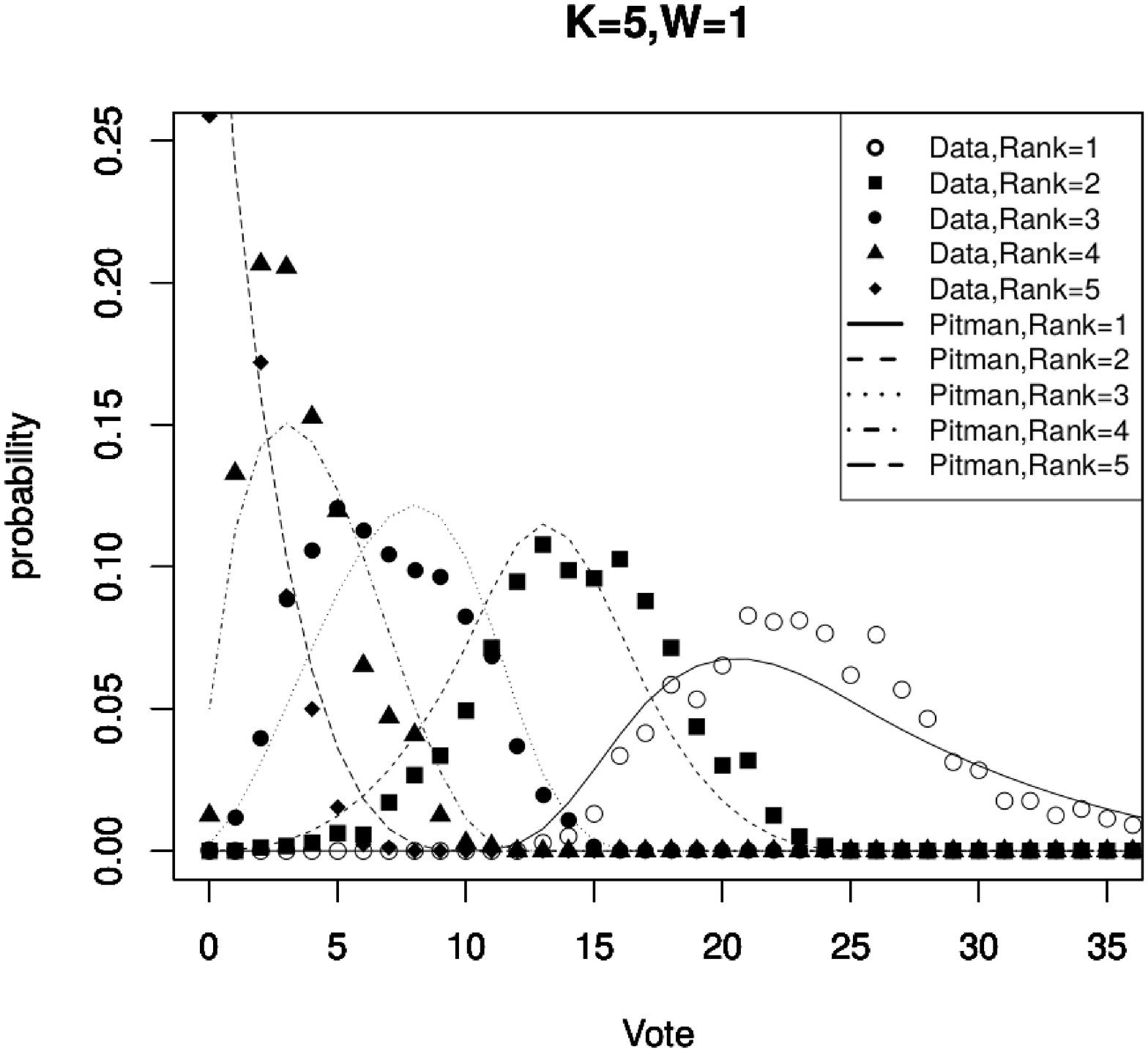} \\
\includegraphics[width=5.5cm]{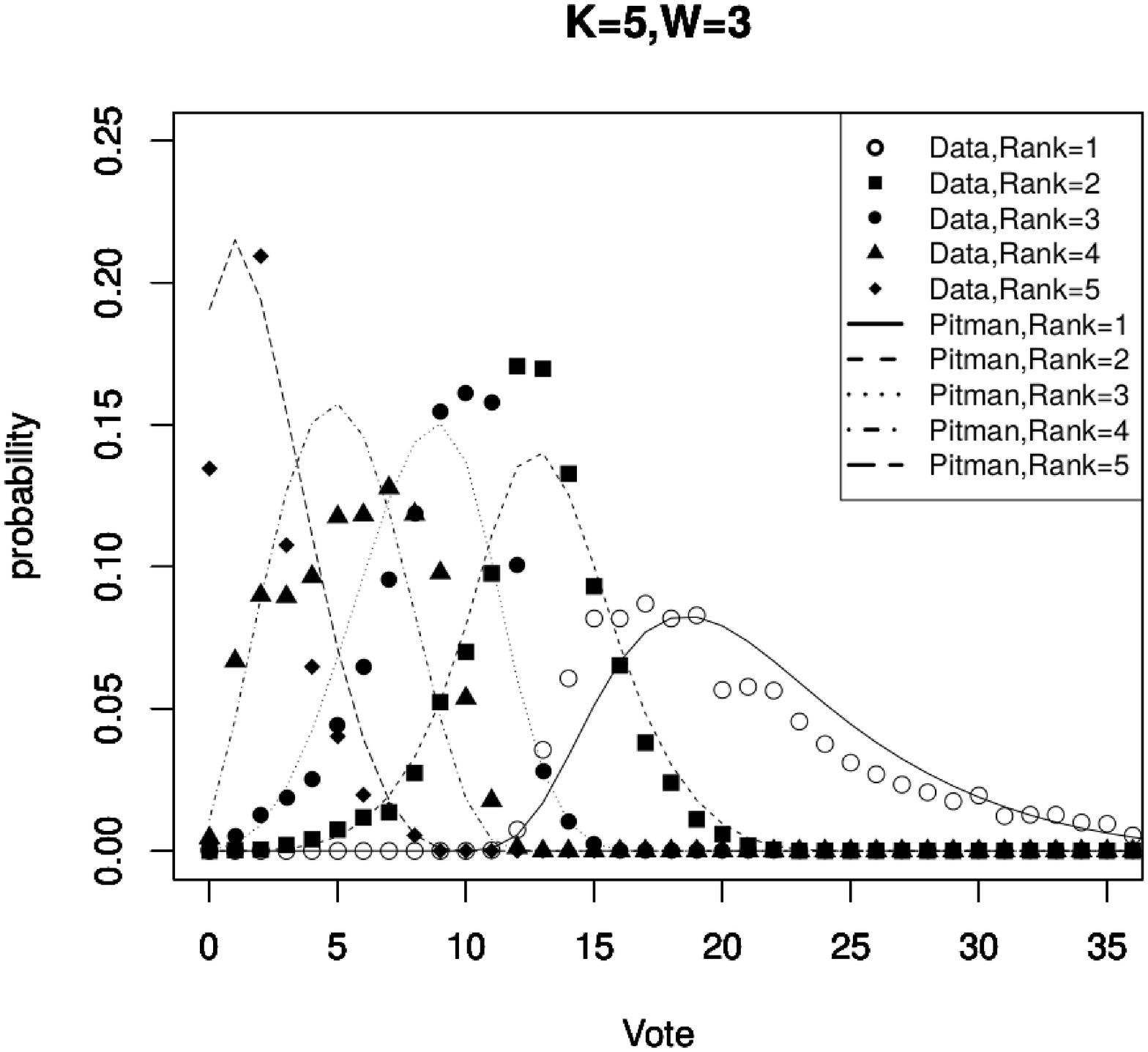} &
\includegraphics[width=5.5cm]{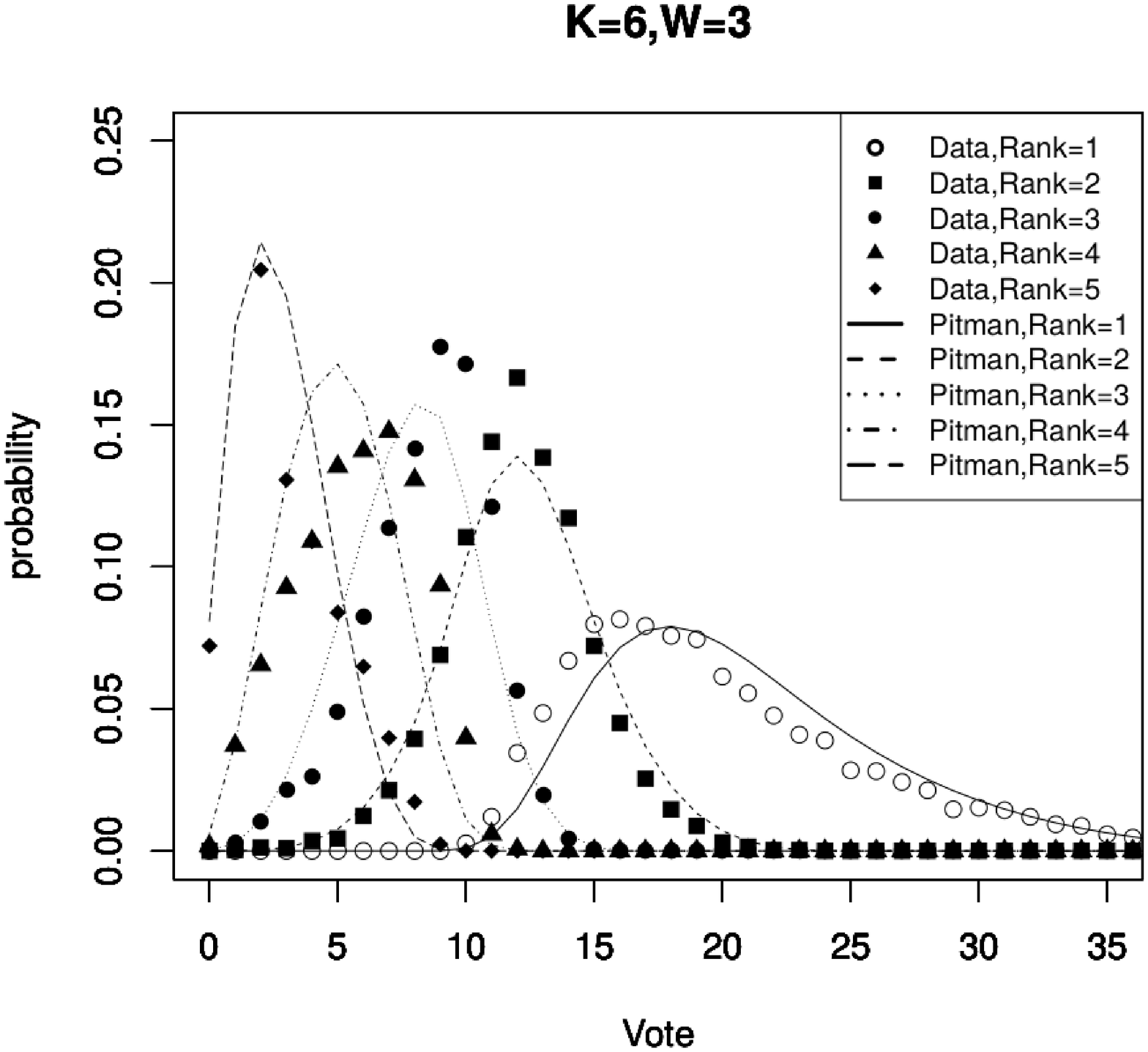} &
\includegraphics[width=5.5cm]{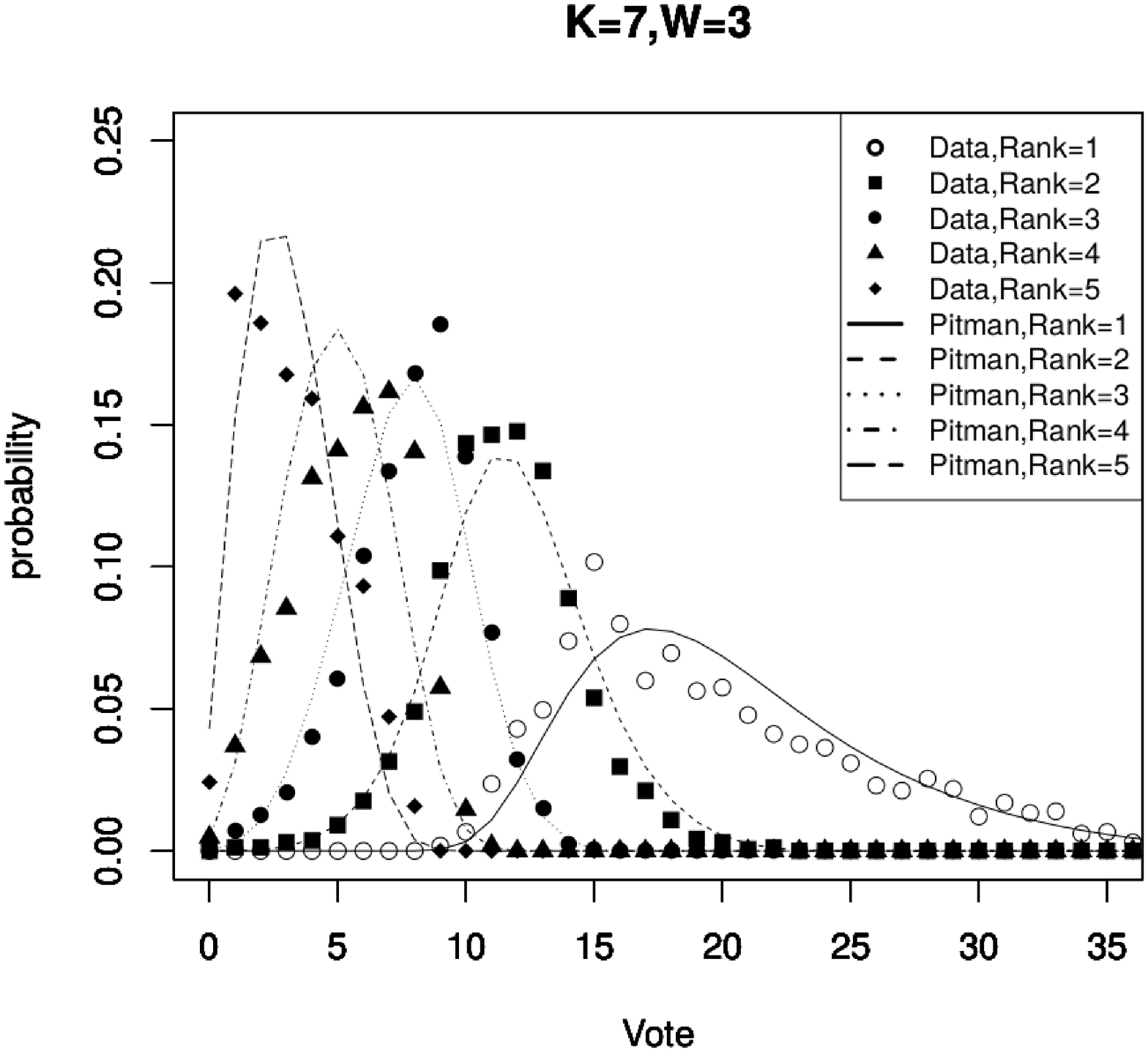} \\
\includegraphics[width=5.5cm]{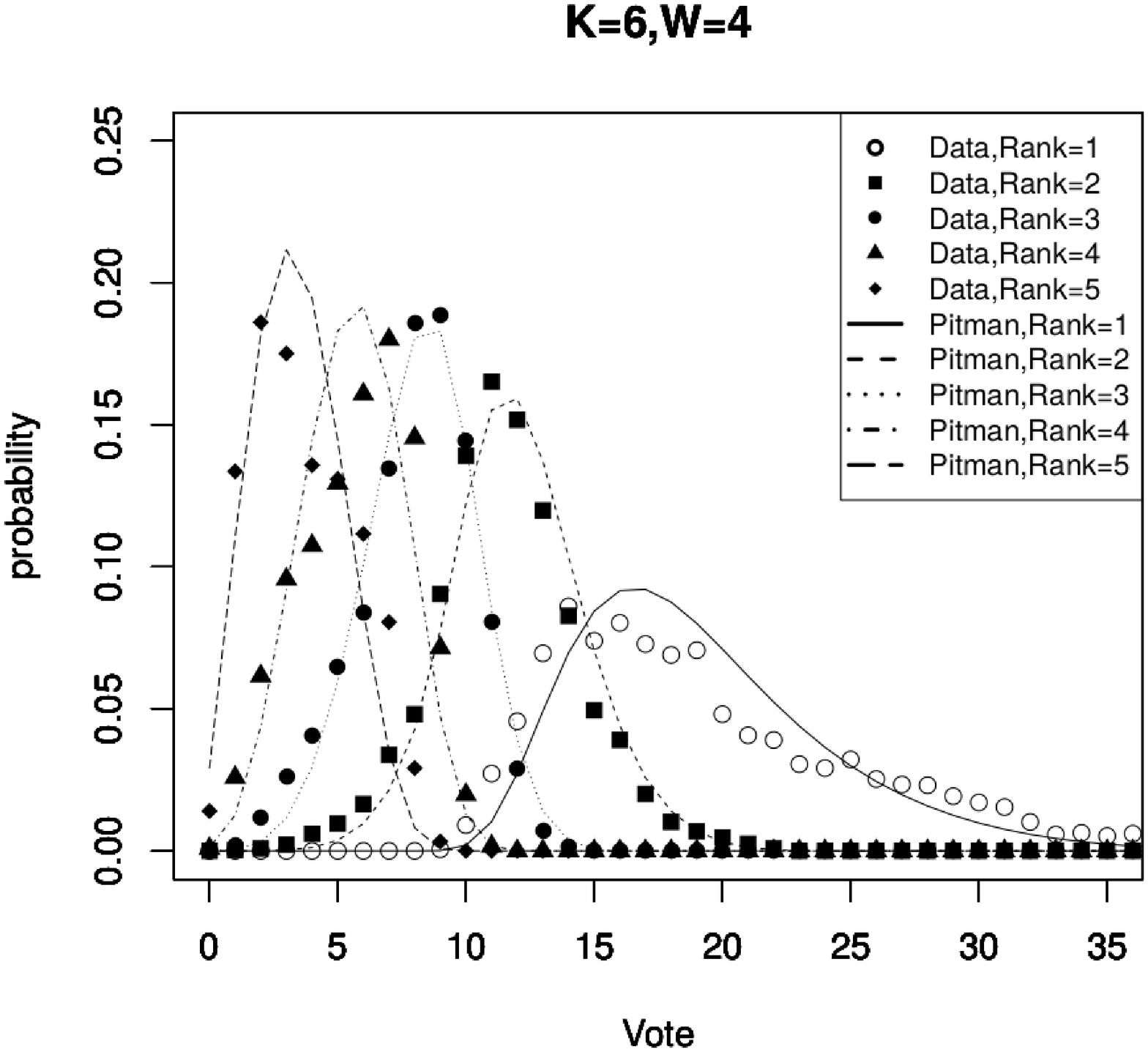} &
\includegraphics[width=5.5cm]{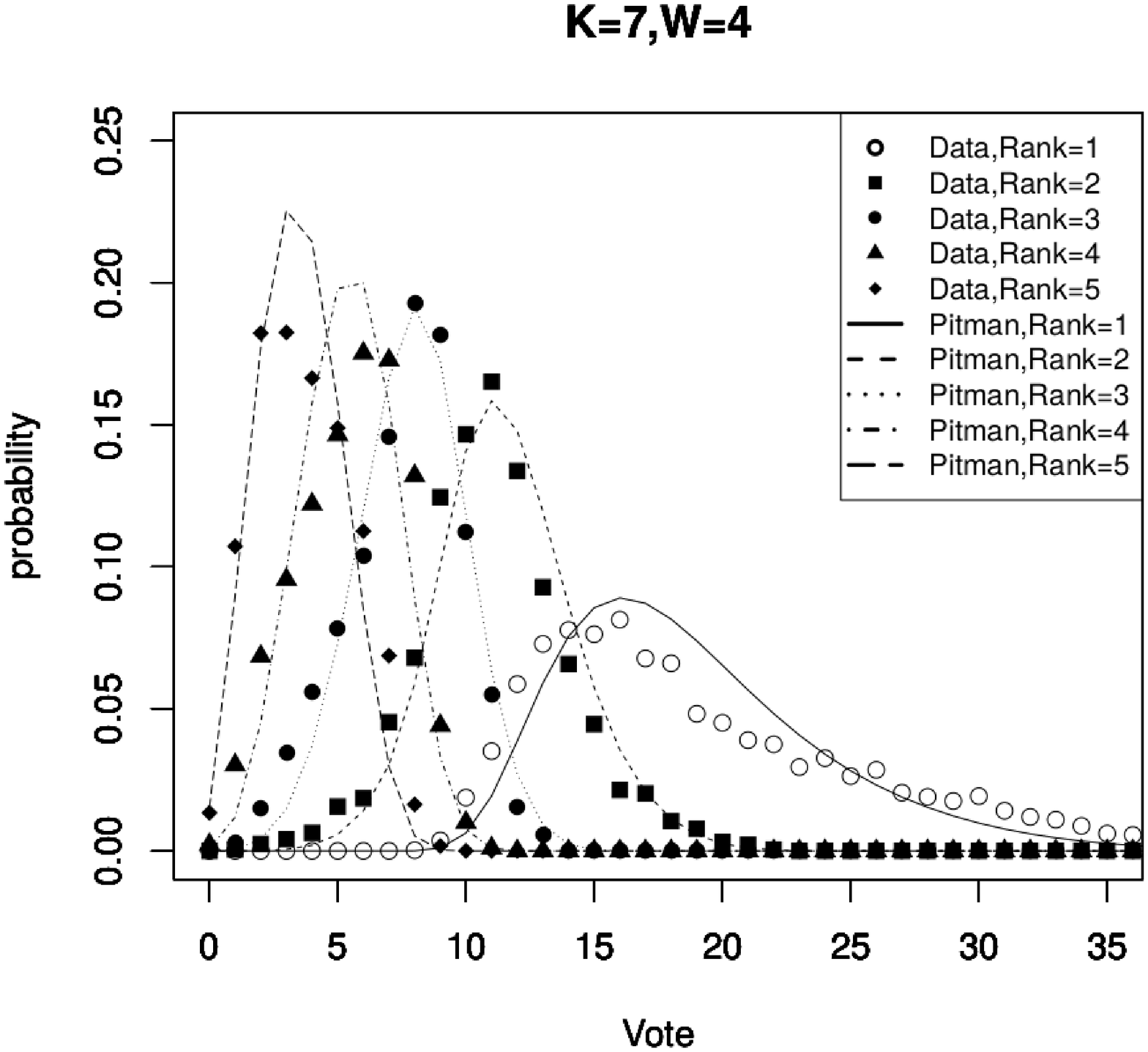} &
\includegraphics[width=5.5cm]{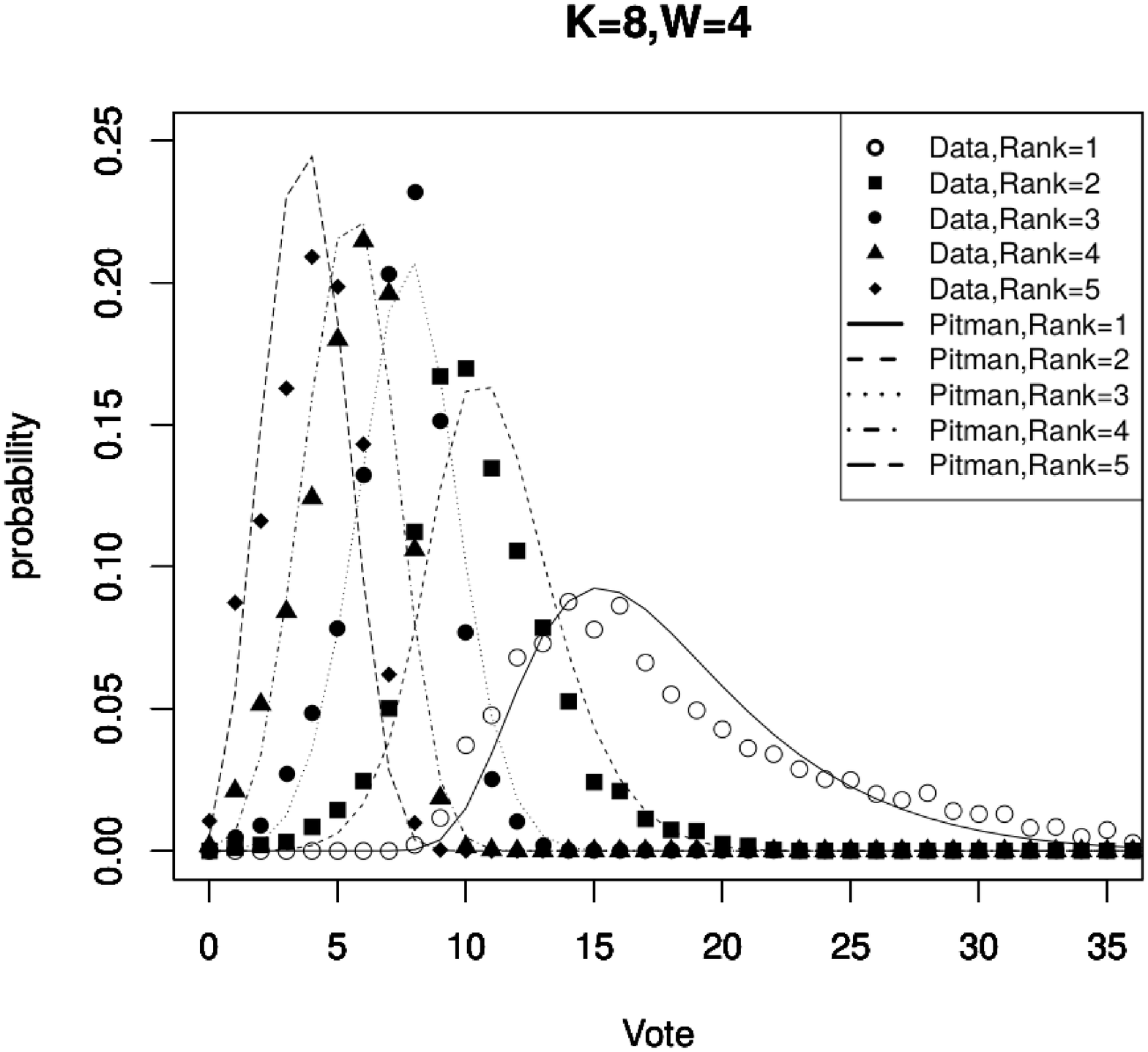} \\
\includegraphics[width=5.5cm]{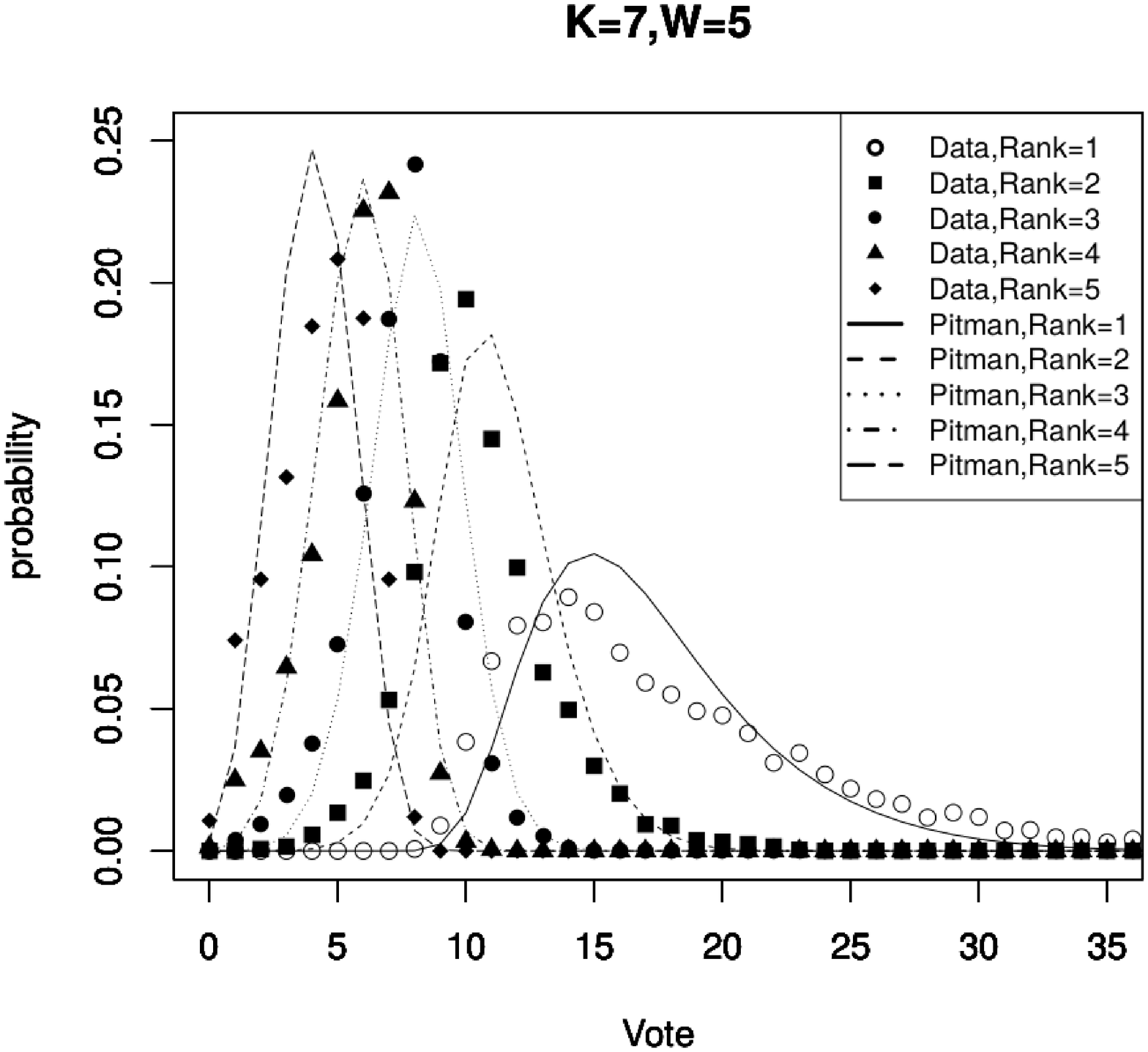} &
\includegraphics[width=5.5cm]{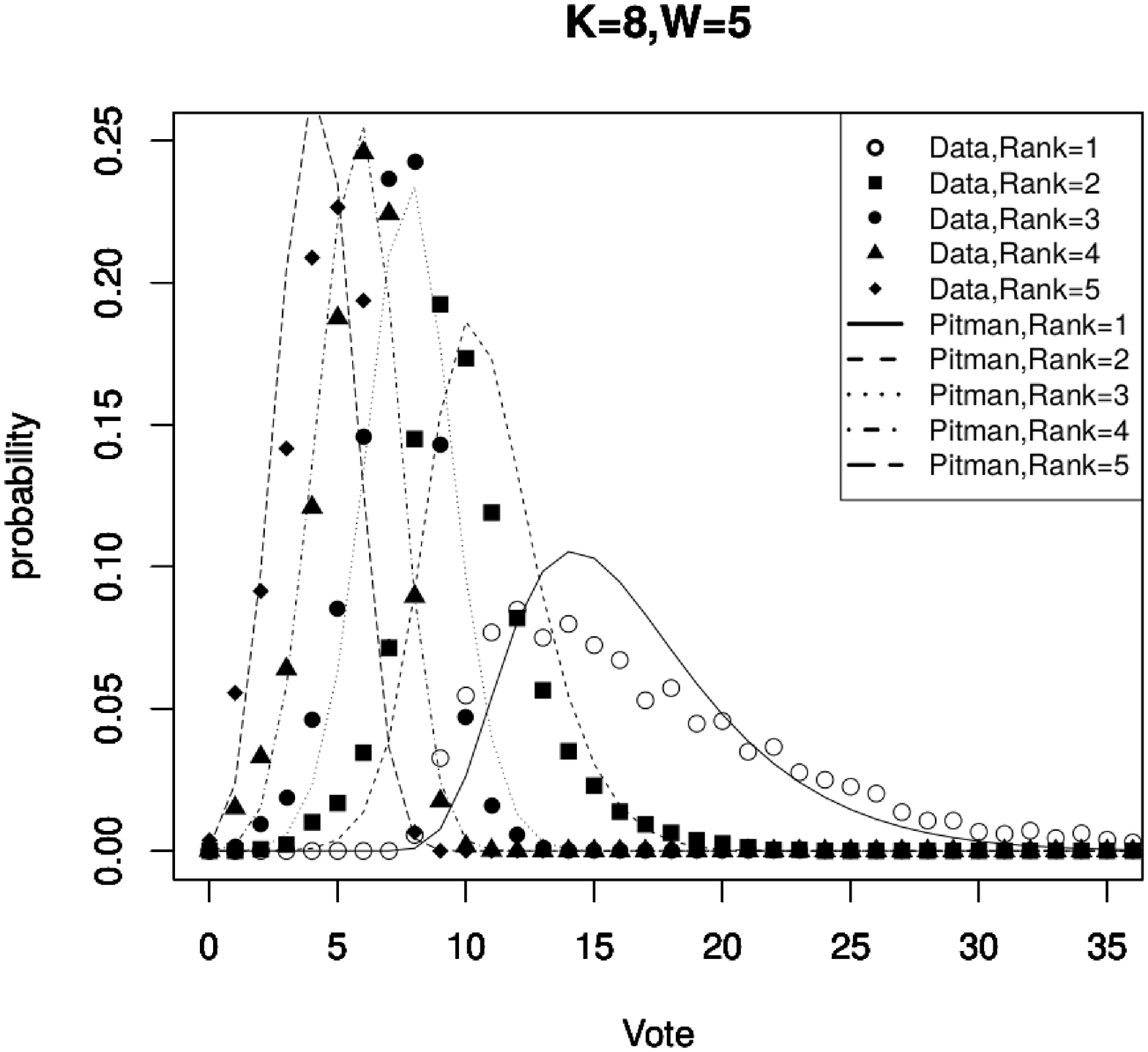} &
\includegraphics[width=5.5cm]{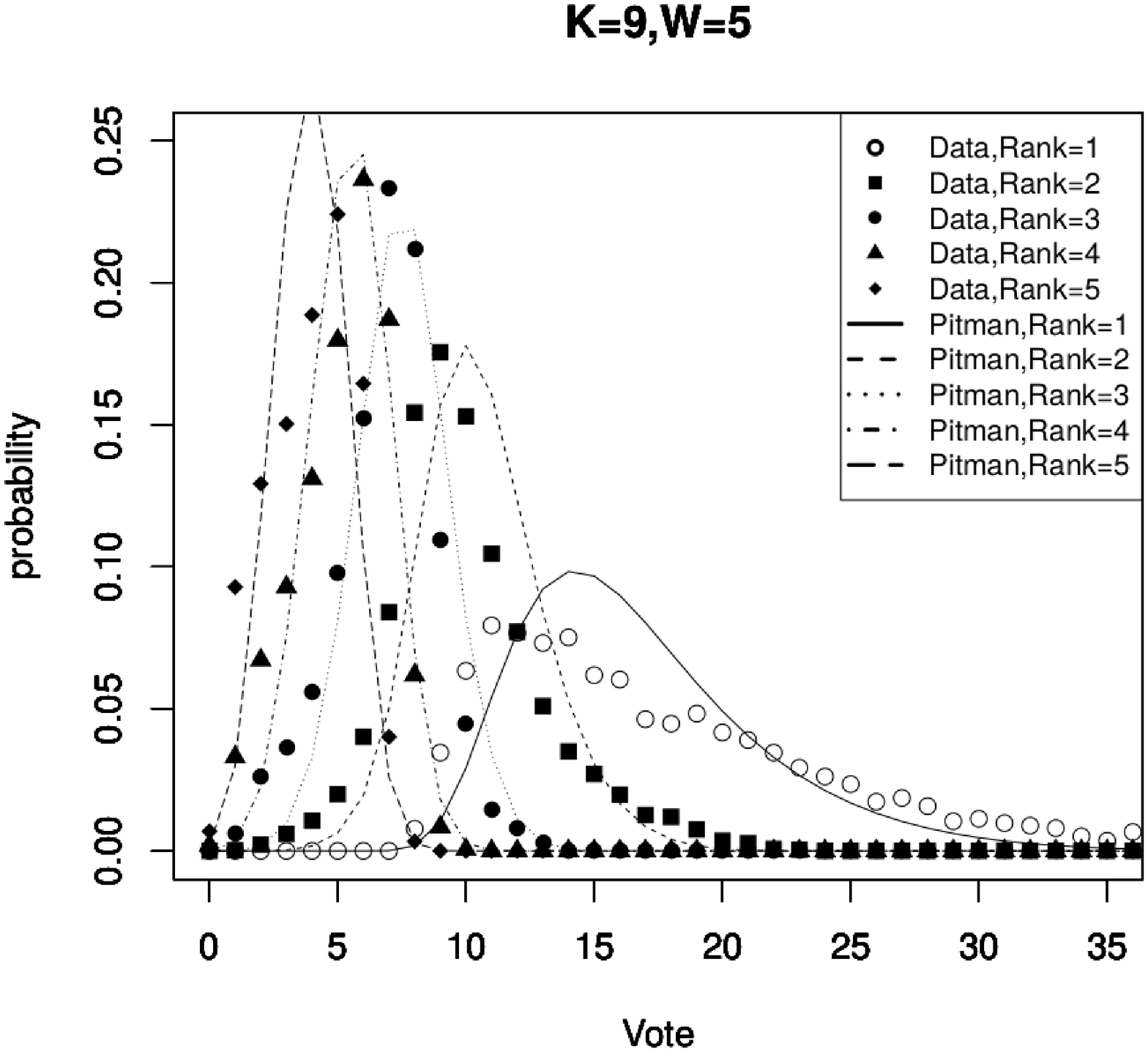} \\
\end{tabular}  
\caption{Plots of the distribution of $c_{k}$ for
  $k\le 5$ and $r=100$. We choose the three cases with the largest
  sample numbers for $W\in \{1,3,4,5\}$.
}
\label{fig:election}
\end{figure*}

To check the fit of the Pitman sampling formula (\ref{pit2}) for
$\theta=-K\alpha$, 
we plot the distribution of $\hat{k}_{i}$ for $r=100$ and 
$i\leq 5$
for the 12 cases in Fig. \ref{fig:election}.
The middle constituency system $W\neq1$ results show a good fit.
On the other hand, this is not the case for the small constituency system,
$W=1$ because we neglected the inhomogeneities of the system, which affect
 the voting results, particularly in the small constituency system.

\section{\label{B} Goodness-of-fit test for 2ch.net data}

Given the observed dataset and the Pitman sampling
formula, we 
test whether the empirical data are truly drawn from the formula.
A standard approach is to perform a goodness-of-fit test,
which generates a p-value that quantifies the plausibility of the hypothesis.
To adopt the procedure, it is necessary
to introduce a measure of the distance between the distribution of
the empirical data and the Pitman sampling formula.
This distance is compared with distance measurements for comparable synthetic
datasets drawn from the Pitman sampling formula, and the p-value is defined as the fraction
of the synthetic distances that are larger than the empirical distance \cite{Clauset:2009}.
If the p-value is rather small, one can reject the Pitman sampling formula as a
plausible fit to the data.

The Pitman sampling formula provides a probability for each decomposition
$\bm{\hat{a}}=(a_{1},a_{2},\cdots,a_{r})$ of integer $r$ as $\sum_{j}ja_{j}=r$.
$K_{r}$ is defined as $K_{r}=\sum_{j}a_{j}$.
We adopt the Kolmogorov--Smirnov (KS) statistic as a
measure quantifying the distance between the Pitman sampling formula
and the empirical data. We explain the procedure below, which is based
on the goodness-of-fit test of power laws \cite{Clauset:2009}.
\begin{itemize}
\item First, we randomly choose $S$ sequences of length $r$ from
  the post data $\{n(t)\},t=1,\cdots,T$. We then randomly draw $S$ integers
  $t_{n},n=1,\cdots,S$ from $\{10^4,\cdots,T-r\}$ and choose
  $S$ sequences as $(n(t_{n}),n(t_{n}+1),\cdots,n(t_{n}+r-1))$.
  
\item We calculate the decomposition $\bm{\hat{a}}_{n}$ for
  the sequence $(n(t_{n}),n(t_{n}+1),\cdots,n(t_{n}+r-1)$
  and obtain the empirical distribution of the decomposition of
  $r$. We then use the maximum likelihood principle to estimate
  $(\theta,\alpha)$. In addition, we estimate the KS statistics for this fit.
  We define the KS statistics as the maximum distance between the
  cumulative distribution functions (CDFs) of the
  data and the fitted model:
  \[
   D=\mbox{max}|S(\bm{\hat{a}})-P(\bm{\hat{a}})|.
  \]
  Here, $P_{r}(\bm{\hat{a}})$ is the CDF of the Pitman sampling formula that best
  fits the data and $S(\bm{\hat{a}})$ is the CDF of the data.
  In calculating the CDFs, we order $\bm{\hat{a}}$
  according to size. 
  
\item We generate the Pitman sampling formula for the distributed synthetic datasets
  with parameters $\theta,\alpha$ equal to those of the distribution that
  best fits the above data. The sample number is $S$.
  We fit the synthetic data to the Pitman
  sampling formula and calculate the KS statistics. We repeat the procedure
  2,500 times and obtain the same number of KS statistics.
  Then, we estimate the 90\% point $KS_{90\%}$ of the distribution of the
  KS statistics. In general, we can count the fraction
  of the synthetic samples whose KS value is larger than the value for
  the empirical data and treat it as the p-value for the empirical data.
  Instead, we calculate the ratio $KS/KS_{90\%}$, where $KS$ in the numerator
  is the KS statistic for the empirical data. If the ratio is greater than
  one, the empirical data's p-value is less than 10\%.
  Thus, we can reject the hypothesis that the empirical
  data obeys the Pitman sampling formula. Undoubtedly, even if the p-value
  is large, it does not
  guarantee that the Pitman sampling formula is the correct distribution
  of the data. Some other model may prove to be a better fit to the data.
  In addition, if the sample size $S$ is too small, the p-value could become
  large.
  We avoid the latter by adopting $S=3\times 10^4$.
  In addition, we estimate the ratio $KS/KS_{90\%}$ because 
  one can use its value as the proxy for the difference between the Pitman sampling
  formula and the empirical data. 
\end{itemize}

\begin{figure}[tbh] 
\includegraphics[width=6cm]{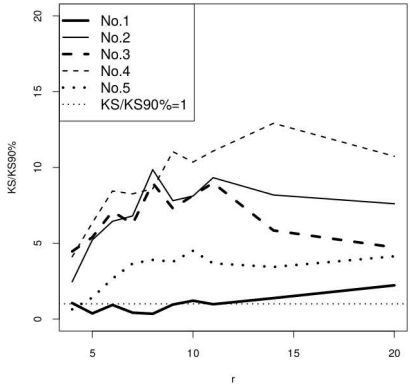} 
\caption{Plot of $KS/KS_{90\%}$ vs. $r$ for the first five boards,
No. 1, 2, 3, 4, and 5. $r\in \{4,5,6,7,8,9,10,11,14,17,20\}$.}
\label{fig:2ch_KS}
\end{figure}

We estimate the ratio $KS/KS_{90\%}$ for the first five boards.
The number of partitions $\hat{a}$ of $r$ for $r=3$
is 3, $3=1+1+1=2+1+3$. We use two parameters as the fit, and 
there are no degrees of freedom.
For $r=4$, there are
five partitions, $4=1+1+1+1=2+1+1=2+2=3+1=4$.
The degree of freedom that fits is $5-2-1=2$.
Here, we adopt an $r$ that is greater than four as $r\in \{4,5,6,7,8,9,10,11,14,17,20\}$.
As $r$ increases, the number of partitions rapidly increases.
For $r=20$, there are 627 partitions, and the remaining degrees of freedom
is 624. 

We illustrate the results in Fig.\ref{fig:2ch_KS}.
We can see that the ratio is less than one for the No. 1 board with
$r=5,6,7,8,9,11$
and the No. 5 board with $r=4$. For the large $r$ for these two boards and
other boards, the Pitman sampling formula is rejected.

\section{\label{aD} Additional information about BBS 2ch.net}
In this appendix, we provide
supplementary information about
2ch.net
and the dataset studied here. 
2ch.net is a collection of multiple BBS and
was founded in May 1999.
According to a NetRating survey, the number of 
2ch.net users in 2009 was 11.7 million.

To obtain post data for 2ch.net,
we chose ten boards for several genres, as shown in Table
\ref{tab:2ch}.
The first five board genres are news, and the other five belong to
various other genres.
The program processes the HTML files
and extracts the number of threads on the boards. 
All threads have a 10-digit ID,  
a post date, and an ID of the user who made the post.
The IDs are randomly assigned by 2ch.net to maintain 
user anonymity.
 If the thread is in sequence with a previous thread,
 we record the ancestor's thread ID.

 The data and R scripts to generate the figures presented in this study
 are available online:
 https://202.24.143.74/2ch.
 The data for the No. $k$ board is denoted by k.csv and comprises
  $(s(t),n(t),id(t)), t=1,\cdots,T$ in three columns. 

\end{document}